\documentclass{jfm}
\usepackage{epstopdf, epsfig}
\usepackage{amsmath,amssymb,amsfonts,bm}
\usepackage[euler]{textgreek}
\usepackage{rotating}
\usepackage{stmaryrd}
\usepackage{xcolor}
\usepackage{cancel}
\usepackage{graphicx}
\usepackage{subcaption}
\usepackage{wrapfig}
\usepackage{multicol}

\newcommand{\Ro}{\mbox{\it Ro}}

\def\begineq{\begin{equation}}
\def\endeq{\end{equation}}

\def\begineqn{\begin{equation*}}
\def\endeqn{\end{equation*}}

\def\beginar{\begin{eqnarray}}
\def\endar{\end{eqnarray}}
\def\beginarn{\begin{eqnarray*}}
\def\endarn{\end{eqnarray*}}

\def\lb{\left ( }
\def\rb{\right ) }
\def\lsq{\left [ }
\def\rsq{\right ] }

\def\ek{E}

\def\ub{\boldsymbol{u}}

\def\hz{\boldsymbol{\widehat \eta}}
\def\haz{\boldsymbol{\widehat z}}
\def\hax{\boldsymbol{\widehat x}}
\def\hay{\boldsymbol{\widehat y}}

\def\hy{{\bf\widehat y}}
\def\hx{{\bf\widehat x}}

\newcommand\Beq{\begin{eqnarray}} 
\newcommand\Eeq{\end{eqnarray}}

\newcommand{\pd}[1]{\partial_{#1}}

\newcommand{\Ra}{\widetilde{Ra}}
\newcommand{\eps}{\varepsilon}

\shorttitle{Parameterized Ekman boundary layers}
\shortauthor{S. Tro, I. Grooms and  K. Julien}

\title{Parameterized Ekman boundary layers on the tilted $f$-plane}

\author{Sara Tro\aff{1}, Ian Grooms\aff{1}, Keith Julien \aff{1}\corresp{\email{julien@colorado.edu}}}

\affiliation{\aff{1}Department of Applied Mathematics, University of Colorado, Boulder, CO 80309, USA}

\begin{document}

\maketitle

\vspace{1.5cc}

\begin{abstract}
\noindent  Rotating convection is considered on the tilted $f$-plane where gravity and rotation are not aligned. For sufficiently large rotation rates, $\textOmega$, the Taylor-Proudman effect results in the gyroscopic alignment of anisotropic columnar structures with the rotation axis giving rise to rapidly varying radial length scales that vanishes as $\textOmega^{-1/3}$ for $\textOmega\rightarrow\infty$. Compounding this phenomenon is the existence of viscous (Ekman) layers adjacent to the impenetrable bounding surfaces that diminish in scale as  $\textOmega^{-1/2}$. In this investigation, these constraints are relaxed upon utilizing a non-orthogonal coordinate representation of the fluid equations where the upright coordinate aligns with rotation axis. This exposes the problem to asymptotic perturbation methods that permit: (i) relaxation of the constraints of gyroscopic alignment; (ii) the filtering of Ekman layers through the uncovering of parameterized velocity pumping boundary conditions; and (iii) the development of reduced quasi-geostrophic systems valid in the limit $\textOmega\rightarrow\infty$.

Linear stability investigations reveal excellent quantitative agreement between results from parameterized or unapproximated mechanical boundary conditions. For no-slip boundaries, it is demonstrated that the associated Ekman pumping dramatically alters convective onset through an enhanced destabilization of large spatial scales.  The range of unstable modes at a fixed thermal forcing is thus significantly extended with a direct dependence on $\textOmega$.  This holds true even for geophysical and astrophysical regimes characterized by extreme values of the non-dimensional Ekman number $\ek$. The nonlinear regime is explored via the global heat and momentum transport of single-mode solutions to the quasi-geostrophic systems which indicate $\mathcal{O}(1)$ changes irrespective of the smallness of $\ek$.

\end{abstract}
\vspace{0.95cc}
\parbox{30cc}{{\it Keywords: B\'enard Convection, Quasi-geostrophic flows,  Boundary layer structure  }
}

\section{Introduction} \label{intro}
Buoyantly driven convection that is constrained by the Coriolis force is a ubiquitous phenomenon occurring within planetary and stellar interiors. It serves as the power source for the generation of large scale magnetic fields \citep{cJ11b,pR13,jmA15}, and may also be the driving mechanism for the observed large scale zonal winds \citep{vasavada2005, yK20} and vortices observed on giant planets \citep{aA2018,lS22}. It is also thought to be an important source of turbulent mixing even within the recently discovered global subsurface oceans of icy moons \citep{kS2019,Bire2020}. Non-dimensional parameters that characterize these geophysical and astrophysical phenomena are extreme. Estimates based on the characteristic flow speed $U$, domain scale $H$, rotation rate $\textOmega$, and kinematic viscosity $\nu$, indicate that the global scale Reynolds number measuring turbulent intensity is large, i.e., 
\begin{subequations}
\label{eqn:ReRoEk}
\Beq
Re_H=\frac{\tau_\nu}{\tau_u}=\frac{U H}{\nu}\gg1.  
\Eeq
with eddy turnover time $\tau_u = H/U$ and viscous diffusion time $\tau_\nu= H^2/\nu$. 
Additionally, the Ekman and bulk Rossby numbers measuring, respectively, the magnitude and constraint of rotation are small, i.e., 
 \Beq 
\label{eqn:geo}
\ek=\frac{\nu}{2\mathrm{\textOmega} H^2}=\frac{\tau_\mathrm{\textOmega}}{\tau_\nu}\ll1, \quad Ro_H=\frac{U}{2\mathrm{\textOmega} H} = Re_H \ek =\frac{\tau_\mathrm{\textOmega}}{\tau_u}\ll 1 
\Eeq 
\end{subequations}
with system rotation time $\tau_\mathrm{\textOmega} = (2\mathrm{\textOmega})^{-1}$.
Also evident from laboratory experiments, numerical simulations, and theory is the existence of
strong spatial anisotropy due to the gyroscopic alignment resulting from the Taylor-Proudman constraint that arises through a leading order geostrophic force balance between the Coriolis and pressure gradient forces \citep{julien2006generalized,kJ07,jmA15}. Anisotropy is quantified by the aspect ratio $A=\ell/H \sim \ek^{1/3} \ll1$ with $\mathcal{O}(\ell)$ non-axial and $\mathcal{O}(H)$ axial eddy length scales.

Equations~(\ref{eqn:ReRoEk}) provide the ordering $\ek\ll Ro\ll 1$ that also implies the relative time ordering $\tau_\mathrm{\textOmega} \ll \tau_u \ll \tau_\nu$. 
As an example, for the Earth’s outer core estimates suggest $Ro=\mathcal{O}(10^{-7})$, $\ek=\mathcal{O}(10^{-15})$, and $Re=\mathcal{O}(10^{8})$ indicating fifteen decades of temporal separation between fast inertial waves that propagate on timescale $\tau_\mathrm{\textOmega}$ and the viscous time $\tau_\nu$ (or seven decades when compared with the eddy turnover time $\tau_u$) \citep{pR13}. These parameters are far beyond the current investigative capabilities of direct numerical simulations (DNS) in both global spherical or local planar domains which remain limited to $\ek\gtrsim \mathcal{O}(10^{-7})$ and $Re_H\lesssim \mathcal{O}(10^4)$. 
This restriction is largely due to the stiffness that arises in simulating the Navier-Stokes equation as a consequence of several factors. Specifically, (i) the aforementioned prohibitive temporal range, (ii) the presence of strong spatial anisotropy, and (iii) the presence of thin viscous (Ekman) boundary layers of $\mathcal{O}(\ek^{1/2}H)$ appearing unconditionally for no-slip boundaries and conditionally for stress-free boundaries when the direction of gravity and axis of rotation are misaligned. In turn, the abatement of these constraint can be achieved by (1) implementing implicit time-stepping treatments for the Coriolis force thus removing the impact of fast inertial waves on the Courant-Friedrich-Levy (CFL) timestepping constraint \citep{Ded2020,bM21}, (2) utilizing an axially-aligned non-orthogonal coordinate system that is scaled anisotropically in horizontal and axial directions \citep{kJ98,aE2023}, and (3)  circumventing  the need to resolve Ekman boundary layers via their parameterization. 
The latter two items (2) and (3) are focal points of the present paper and explored within the configuration for rotating Rayleigh-B\'enard convection (RRBC) within the tilted $f$-plane approximation located at an arbitrary co-latitude $\vartheta_f$.  

For upright RRBC, \citet{pN65,wH71} and \citet{homsy1971} first established the quantitative difference between the critical onset of convection in the presence of no-slip and stress-free boundaries as an $\mathcal{O}(Ek^{1/6})$ asymptotic correction. The existence of a boundary condition parameterizing the effect of Ekman pumping for this case was first uncovered by \citet{jCKMSV16}. However, to-date, a full exploration of the impact of Ekman pumping on marginal onset and the asymptotic robustness of parameterized boundary conditions at finite $(\ek,Ro)$ has yet to be performed. Moreover, these open questions extend to the more geophysically relevant RRBC on the tilted $f$-plane. Here, it is also known  that Ekman boundary layers also exist in the presence of stress-free mechanical boundaries \citep{kJ98}. However, to our knowledge, irrespective of the type of mechanical boundary condition, the precise nature of \emph{parameterized} boundary conditions on the tilted $f$-plane remains open and uncovered in this paper for the first time. For no-slip boundaries, it is demonstrated that Ekman pumping strongly destabilizes the onset of convection at large scales to an extent that the range of unstable wavenumbers is greatly extended. In the nonlinear regime, it is found that pumping results in a net transport of heat due to a direct correlation between thermal and vertical velocity fluctuations that strongly enhances the global heat flux. For stress-free boundaries, it is demonstrated that despite the existence of Ekman boundary layers, no net heat transport occurs due to a $90^\circ$ phase-lag between thermal and vertical velocity fluctuations.

The organization of this paper is as follows. In section~\ref{sec:formulation}, the RRBC problem on the tilted $f$-plane is formulated with the incompressible Navier-Stokes equations (iNSE) along with its asymptotic reduction to the low-$Ro$ quasi-geostrophic equations (QG-RBC) that constitute a foundation for a point of comparison throughout for all results presented.
For analytic and numerical advancement, a non-orthogonal coordinate representation is pursued where the upright coordinate is taken to be the axis of rotation as opposed to the vertical coordinate of gravity. In section~\ref{sec:boundary}, a matched asymptotic analysis is performed on the tilted $f$-plane establishing the existence of three regions: an inner Ekman boundary layer (section~\ref{sec:Ekman}), a middle thermal wind layer (section~\ref{sec:TW}), and an outer or interior region (section~\ref{sec:interiorandpumping}). It is demonstrated  that the Ekman boundary layer dynamics is captured by the classic fourth-order linear ODE system \citep{hG69} but with the axial direction serving as the boundary coordinate. This generic result holds irrespective of the selected co-latitude away from the equator. 
Parameterized boundary conditions determined entirely in terms of outer region variables are presented in section~\ref{sec:interiorandpumping} for no-slip and stress-free mechanical boundaries. Extension of the QG-RBC to incorporate parameterized boundary conditions is formulated in section~\ref{sec:CQGRBC} as the Composite QG-RBC. Analytic and numerical results for the linear stability problem for the marginal onset of convection in the quasi-geostrophic limit is discussed in section~\ref{sec:linstab} along with a hypothesis of its sensitivity to Ekman pumping and predictions of a critical wavenumber at which it achieves dominance and departs quantitatively from the stress-free case (section~\ref{sec:linstabdep}). Section~\ref{sec:singlemode} formulates the problem computing for fully-nonlinear exact single-mode solutions to the QG-RBC and CQG-RBC permitting an analysis of the impact of Ekman pumping into the nonlinear regime. Discussion and concluding remarks are found in Section~\ref{sec:discussion}.


\section{Formulation and Preliminaries} \label{sec:formulation}
\begin{figure}
    \centering
    \includegraphics[width = .7\linewidth]{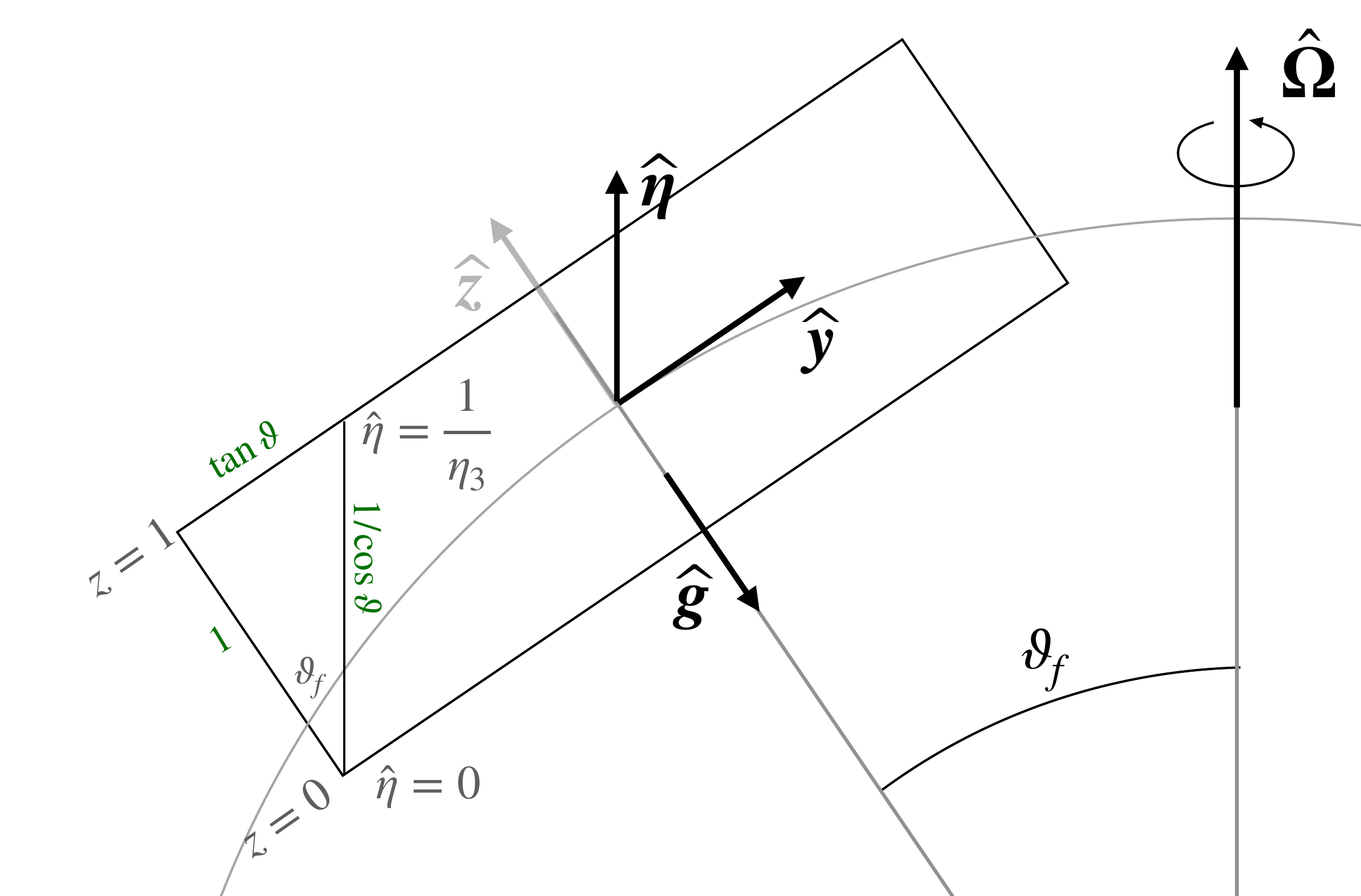}
    \caption{\small{Slice of the local $f$-plane domain along a meridian. Spatial coordinates are non-dimensionalized with respect to layer depth $H$. Here, $\hax$ represents the zonal direction (out of the page), $\hay$, the meridional direction, and $\hz =\eta_2 \hay +\eta_3 \haz $ the local axis of rotation. It follows that a box ranging from $z=0$ to $z=1$ has $\hat{\eta}$ values ranging from $0$ to $1/\eta_3$ with $\eta_3=\cos\vartheta_f$ and where $\vartheta_f$ denotes the co-latitude. }}
    \label{fig:nonorthog_coords}
\end{figure}
\subsection{Incompressible Navier-Stokes Equations}
\label{sec:govequations}
We consider thermal convection on the tilted $f$-plane in the classical Rayleigh-B\'enard configuration, i.e., in a horizontal plane layer of depth $H$ heated from below and cooled from above rotating with a constant angular velocity $\textOmega$ relative to the rotation axis $\hat{\boldsymbol{\Omega}}$. The plane layer is considered to be tangent to a spherical shell at a reference co-latitude $\vartheta_f$ 
(see figure~\ref{fig:nonorthog_coords}).  In the rotating reference frame, fluid motions are assumed incompressible and governed by the Navier Stokes system of equations (iNSE) under the Boussinesq approximation. In non-dimensional form  
\begin{subequations}
\label{eq:NSE}
\beginar
\pd{t}\ub +\ub \cdot \nabla \ub +\frac{1}{\eta_3\Ro}\ \hz\times \ub + Eu \nabla p &=& \frac{1}{Re} \nabla^2 \ub +\Gamma 
 \theta \haz,\\
\pd{t}\theta +\ub \cdot \nabla\theta -A\, \haz\cdot\ub &=& \frac{1}{Pe}\nabla^2 \theta,\\
\nabla \cdot \ub &=& 0,
\endar
\end{subequations}
where $\ub, p, \theta$ are respectively the velocity, pressure and convecting temperature fields. The iNSE is non-dimensionalized by characteristic velocity scale $U$, horizontal length scale $\ell$, 
advective timescale $\ell/U$, pressure scale $P$ and temperature difference $\Delta T$. This results in the appearance of non-dimensional parameters given by 
\begin{equation}
Ro = \frac{U}{2\textOmega \eta_3 \ell}, \quad \Gamma = \frac{g\alpha \Delta T \ell}{U^2},\quad Eu = \frac{P}{\rho_0 U^2}, \quad Re = \frac{U\ell}{\nu}, \quad Pe = \frac{U\ell}{\kappa }, \quad A=\frac{\ell}{H}. 
\end{equation}
Respectively, the Rossby, buoyancy, Euler, Reynolds, Pecl\'et, and aspect ratio numbers with  
 $g$ the acceleration due to gravity, $\alpha$ the coefficient of thermal expansion, $\rho_0$ the constant fluid density, $\nu$ the kinematic viscosity, and $\kappa$ the thermal diffusivity. Importantly, we note that the Rossby number is based on the Coriolis parameter $2\Omega\eta_3$, and henceforth interpreted as the colatitudinal Rossby number. 

The local coordinate system for the iNSE may be defined by Cartesian orthogonal unit vectors $(\hax, \hay, \haz)$ pointing in the zonal (east-west), meridional (north-south), and radial (vertical) directions, respectively. The local velocity field 
is given by 
$\ub = u \hax+ v\hay +w \haz$ and temperature field by $T =T_b+\theta$ where $T_b= 1- A z$
denotes the destabilizing background temperature profile with gradient $-A$. The $f$-plane approximation assumes the constant local rotation vector can be decomposed locally according to $\hz = \eta_2\hay +\eta_3 \haz$ with
\begin{equation}
    \eta_2 = \sin(\vartheta_f), \quad \eta_3 = \cos(\vartheta_f), \quad \gamma =\eta_2/\eta_3= \tan(\vartheta_f).
\end{equation}
 For rotionally constrained thermal convection it has been established that $A\sim Ro\ll1$ characterising the columnar spatial anisotropy of thermal convection \citep{julien2006generalized,AHJPRR20}. Upon selection of a diffusive velocity scale $U=\nu/\ell$ as a reference velocity, where $\ell=Ek^{1/3} H$ is the diffusive length scale, we obtain 
\begin{equation}
 Ro=Ek^{1/3}\equiv\varepsilon\ll1,\quad\mathrm{where}\quad    Ek = \frac{\nu}{2(\textOmega \eta_3)H^2}
\end{equation}
is the co-latitudinal Ekman number. 
This yields the canonical representation of non-dimensional parameters for rotating Rayleigh-B\'enard convection (RRBC)
\begin{equation}
   Re = 1, \quad 
   Pe = \sigma, \quad
   \Gamma = \frac{Ra\eps^3}{\sigma},\quad 
   Eu = \eps^{-2},\quad
    A = \eps,
\end{equation}
where $\sigma = \nu/\kappa $ is the Prandtl number, assumed $\mathcal{O}(1)$, and $Ra = g\alpha \Delta TH^3/(\nu \kappa)$ is the thermal Rayleigh number.  

With these $\eps$-dependent distinguished limits, a leading order geostrophic balance  
\beginar
\label{eqn:geos}
\varepsilon^{-1} \hz\times\ub+\varepsilon^{-2} \nabla p \approx \boldsymbol{0},
\endar
with $\nabla p\sim\mathcal{O}(\varepsilon)$ is observed at $\mathcal{O}(\varepsilon^{-1})$ in equation (\ref{eq:NSE}a) of the iNSE. 
Along with incompressibility (\ref{eq:NSE}c), the Taylor-Proudman constraint 
\beginar
\hz\cdot\nabla(\ub,p)\approx 0
\endar
follows from \eqref{eqn:geos} and operates axially on small $\mathcal{O(\ell})$ dimensional length scales \citep{julien2006generalized}. Given $\ell \ll H$, axial modulations of $\mathcal{O}(H)$ spatial scales are permitted without violation of the Taylor-Proudman constraint. 
Following  \citet{julien2006generalized}, it is therefore convenient to pose the iNSE (\ref{eq:NSE}) in the non-orthogonal coordinate system defined by the unit directions $(\hax, \hay, \hz)$ and where $\ub=u\hx+(v-\gamma w)\hy + w/\eta_3 \hz$.
For RRBC, the non-dimensional radial coordinate $z$ ranging from $0$ to $1$ (in units of $H$) implies an axial coordinate $\tilde{\eta}$ ranging from $0$ to $1/\eta_3$, as shown in figure \ref{fig:nonorthog_coords}. We find it convenient to rescale $\tilde{\eta}$ in the $\hz$ direction as $\eta= \eta_3 \tilde{\eta}$ such that $\eta\in(0,1)$. 
All fluid fields are now consider as  functions of non-orthogonal coordinates $(x,y,\eta)$ such that the small-scale Taylor-Proudman constraint becomes $\partial_{\eta} (\ub,p)=o(1)$.\footnote{Throughout this paper $f(x)= \mathcal{O}(\delta)$ implies $\limsup_{\delta\rightarrow 0}{\lVert f(x)\rVert/ \delta =c  <\infty}$ and $f(x)= o(\delta)$ implies $\limsup_{\delta\rightarrow 0}{\lVert f(x)\rVert/ \delta  }=0$.}  We thus invoke modulation on larger axial scales, i.e., the layer depth scale (interpreted in units of $\ell$) with $\pd{\eta} \mapsto \eps\pd{\Omega}$ where $\Omega\in(0,1)$ is the rescaled axial coordinate.

Upon decomposition of fluid variables into mean horizontally averaged (overbarred)  and fluctuating (primed)  components, i.e.,  $f=\overline{f}+f'$, geostrophy requires $\nabla p' = \mathcal{O}(\varepsilon)$ such that $p'\mapsto \eps p'$, and from $\Gamma$ the subdominance of buoyancy requires $Ra \vartheta'/\sigma =o(\varepsilon^{-4})$. The leading order temperature fluctuating equations requires $\theta' \mapsto \eps \theta'$ such that $Ra=o(\varepsilon^{-5})$.
The projection of momentum equation (\ref{eq:NSE}a) onto unit bases 
$\{ \boldsymbol{\widehat{g}}_j\}\equiv (\hax, \hay, \hz) $ gives
{\footnotesize
\begin{subequations}
\label{eq:covariantbasisequationsrescaledfull}
\beginar
\lb \pd{t}  + \ub \cdot \nabla\rb u -\frac{1}{\eps}\lb v-\gamma w \rb +\frac{1}{\eps}\pd{x}p' &=& \nabla^2 u,\\
\lb \pd{t}+  \ub \cdot \nabla\rb \lb v-\gamma w\rb +\frac{1}{\eps} \frac{1}{\eta_3^2} \lb u+\pd{y}p'\rb - \gamma \pd{\Omega}p' &=&  \nabla^2 \lb v-\gamma w\rb -\frac{\gamma \Ra}{ \sigma} \theta',\\
\lb \pd{t}  + \ub \cdot \nabla \rb w -\frac{\gamma }{\eps} \lb u + \pd{y} p'\rb +\pd{\Omega}p' & = &  \nabla^2 w + \frac{\Ra }{ \sigma} \theta',\\
\lb \pd{t} +\ub \cdot \nabla\rb \theta - \eps w &=& \frac{1}{\sigma}\nabla^2 \theta,\\
\pd{x}u +\pd{y} \lb v-\gamma w\rb +\eps \pd{\Omega}w &=& 0
\endar
\end{subequations}
where $\Ra\equiv Ra \eps^{4}$ is the reduced co-latitudinal Rayleigh number.
We note that this projection is achieved via  application of the dot product of the dual coordinates
 vector bases $\boldsymbol{g}^{\hx}=\hx$,  $\boldsymbol{g}^{\hy}=\hy-\gamma \haz$, $\boldsymbol{g}^{\hz}=\haz/\eta_3$ with orthogonality property
$\boldsymbol{\widehat{g}}^i\cdot\boldsymbol{\widehat{g}}_j=\delta^i_{j}$ where $\delta^i_{j}$ is the Kronecker delta function. The advection and diffusion operators are given by
}
\begin{equation}
    \ub \cdot \nabla = u \pd{x} +\lb v- \gamma w\rb \pd{y} + \eps w \pd{\Omega}, \quad \nabla^2 = \pd{x}^2 + \frac{1}{\eta_3^2}\pd{y}^2 -2\eps \gamma \pd{y}\pd{\Omega}+ \eps^2\pd{\Omega}^2.
\end{equation}
We find a subdominant mean velocity field $\overline{\ub}=\mathcal{O}(\eps^2)$ such that to leading order $\ub\approx\ub'$. This results in a leading order mean hydrostatic balance $\pd{\Omega} \overline{p} \approx (\Ra/\sigma)\; \overline{\theta}$.

The iNSE system \eqref{eq:covariantbasisequationsrescaledfull} is accompanied with boundary conditions. We assume periodic boundary conditions in the horizontal direction. We also consider here impenetrable, fixed temperature boundary conditions
\begin{equation}
    w = \theta = 0,\quad \mbox{on}\ \Omega = 0,1,
    \label{eq:wthetaBCs}
\end{equation}
along with either no-slip ($NS$) or stress-free ($SF$) mechanical boundary conditions 
\begin{subequations}
\label{eq:noslipandstressfreeBCs}
\beginar
NS:&\quad (u,v) &= 0,\quad \mbox{on}\ \Omega = 0,1,\\
SF:&\quad  \haz\cdot\nabla(u,v)  = \lb \eps \pd{\Omega}-\gamma\pd{y} \rb (u,v) &= 0,\quad \mbox{on}\ \Omega = 0,1.
\endar
\end{subequations}

\subsection{Reduced quasi-geostrophic  model}
\label{sec:QGreduced}
Of particular utility as a point of comparison is the reduction of the iNSE (\ref{eq:covariantbasisequationsrescaledfull}) to Quasi-Geostrophic Rayleigh-B\'enard Convection model (QG-RBC) of \cite{julien2006generalized}  in the limit of rapid rotation, $\eps  
\to 0$ \citep[see also][]{aE2023}. 
The model is useful for obtaining analytic asymptotic results that serve as a benchmark for results deduced from the iNSE. 
Substitution of the asymptotic expansion
\begin{equation}
    \boldsymbol{v} = \boldsymbol{v}_0 +\eps \boldsymbol{v}_1 +\eps^2 \boldsymbol{v}_2 + \eps^3 \boldsymbol{v}_3 +...,
\end{equation}
where $\boldsymbol{v} = (u,v,w,p,\theta)^T$, into the system (\ref{eq:covariantbasisequationsrescaledfull}) results in geostrophic balance \eqref{eqn:geos} at leading order. Defining a geostrophic streamfunction $\psi_0$ and setting 
\begin{equation}
   u_0 = -\pd{y}\psi_0,\quad v_0-\gamma w_0 = \pd{x} \psi_0,\quad p_0 = \psi_0,
\end{equation}
solves the problem at leading order. At the next highest order, the resulting nonhomogeneous PDE system has associated solvability conditions that imply  
the reduced quasi-geostrophic model for rotating RBC on the tilted $f$-plane (QG-RBC), namely, 
\begin{subequations}
\label{eq:QGRBC}
    \beginar
\pd{t}\nabla_\perp^2\psi_0 +J\left[\psi_0,\nabla_\perp^2\psi_0\right] -\pd{\Omega}W_0 +\gamma \frac{\Ra}{\sigma} \pd{x}\theta'_1  &=& \nabla_\perp^2 \nabla_\perp^2 \psi_0,\\
 \pd{t} W_0 +J\left[\psi_0,W_0\right]  +\pd{\Omega} \psi_0 &=& \nabla_\perp^2 W_0 +\frac{\Ra}{\sigma} \theta'_1,\\
 \pd{t} \theta'_1 +J\left[\psi_0,\theta'_1\right] + w_0 \left (\partial_{\Omega} \overline{\Theta}_0 - 1\right) &=&\frac{1}{\sigma}\nabla_\perp^2 \theta'_1,\\
 \partial_\Omega \lb\overline{w_0\theta'_1} \rb &=& \frac{1}{\sigma} \partial_{\Omega\Omega} \overline{\Theta}_0
    \endar
\end{subequations}
where $\nabla_\perp^2 = \pd{x}^2 + \eta_3^{-2}\pd{y}^2$, $W_0 = \eta_3^{-2} w_0 + \gamma \pd{x}\psi_0$ and  temperature is decomposed into leading order mean and fluctuating components, i.e.,  $\theta= \overline{\Theta}_0 + \varepsilon \theta'_1$ such that $\overline{\theta'_1}=0$. The nonlinear terms have been written in terms of the Jacobian advection operator,
\begin{equation}
    \ub_{0\perp} \cdot \nabla_\perp = u_0\pd{x} +(v_0-\gamma w_0)\pd{y} = \pd{x}\psi_0\pd{y} - \pd{y}\psi_0\pd{x}  = J\left[\psi_0,\cdot \right].
\end{equation}
The QG-RBC is fourth-order in $\Omega$, thus for closure, it is accompanied by the boundary conditions (\ref{eq:wthetaBCs}) applied to $w_0$ and $\overline{\Theta}_0$ on $\Omega=0,1$. From (\ref{eq:QGRBC}c), the variance $\overline{\theta^{\prime2}_1}$ satisfies the equation $\pd{t}\overline{\theta^{\prime2}_1}=-\sigma^{-1}\overline{\vert \nabla_\perp\theta'_1\vert^2}$ implying $\lim_{t\rightarrow\infty}\overline{\theta^{\prime2}_1}=0$. Thus irrespective of the thermal boundary condition on $\overline{\Theta}$ the criteria $\theta'_1=0$ on $\Omega=(0,1)$ is automatically satisfied if its initial value satisfies this boundary condition.

The QG-RBC are valid provided $Ro\ll1$ which holds for $\Ra = o(\eps^{-1})$, or equivalently, $Ra = o(Ek^{-5/3})$ \citep{mS06,julien2006generalized,kJ12,jCKMSV16}.
By definition, given $\ub^*=(\nu/\ell) \ub$ dimensionally, then 
\beginar
\label{eqn:Roc}
Ro = \frac{\nu}{2\mathrm{\textOmega}\eta_3\ell^2} \lVert \ub \rVert = \eps \lVert \ub \rVert.
\endar
It follows $\lVert \ub \rVert\sim\lVert \psi_0 \rVert\sim\lVert \zeta_0 \rVert =o( \eps^{-1})$
for rotational constraint, where $\zeta_0 = \pd{x}v_0-\pd{y}u_0$ is the radial vorticity. We note that the solutions to the QG-RBC can be generally viewed asymptotically as an outer solutions because they do not automatically satisfy the mechanical no-slip or stress-free boundary conditions \eqref{eq:noslipandstressfreeBCs}. This requires boundary layer corrections via matched asymptotics that are discussed in the next section.

\section{Boundary layers}
\label{sec:boundary}
While the interior of the domain for the iNSE system \eqref{eq:covariantbasisequationsrescaledfull} 
is dominated by a leading order geostrophic balance, standard choices of mechanical boundary conditions are incompatible with this balance on the tilted $f$-plane. Ekman boundary layers, where the dominant force balance transitions from geostrophy to include viscous stresses, are thus generated at the top and bottom of the domain \citep{hG69,kJ98}. The QG-RBC system \eqref{eq:QGRBC} filters Ekman layers and thus may be evolved solely with the knowledge that the boundaries are impenetrable and fixed temperature. This is consistent with the observation that the QG-RBC is fourth-order in $\Omega$. However, this side-lines any assessment of the impact of mechanical boundaries.

It is well-established for the upright case ($\vartheta_f = 0^\circ$) that impenetrable no-slip boundaries generate Ekman layers while stress-free boundary conditions do not \citep{kJ98}. Here, we generalize the theory to non-zero tilt angles (co-latitudes) where we find, \textit{a posterori}, all mechanical boundary conditions generate Ekman layers. The ultimate objective of this section is to uncover the parameterized boundary conditions  in terms the interior fluid variables that characterize the dynamical impact of an Ekman layer and thereby alleviate the need to resolve it numerically. These are  often referred to as pumping conditions (generically taken to capture the action of both pumping and suction). We demonstrate in this section that away from the equatorial region (i.e., for $\gamma =o(\eps^{-1/2})$), the system of boundary layer equations valid in the Ekman layer have the classical ODE form for the upright case consisting of a fourth order linear operator in space albeit now operating in the axial direction.

The boundary layer theory is formulated by decomposing the fluid variables into an outer component  (for the geostrophic interior) and inner components at the upper and lower boundaries located at  $\Omega=0,1$ (for the Ekman boundary layers). \cite{jCKMSV16} have established that for no-slip boundaries the presence of an Ekman boundary layer also drives a thermal wind layer (a middle boundary layer region), a required thermal response to satisfy the thermal boundary condition $\theta^\prime=0$ on $\Omega= 0,1$. We establish in  section~\ref{sec:TW} that no such thermal wind layer is required in the presence of stress-free boundaries, thus to leading order the fixed temperature boundary conditions are automatically satisfied without need of a boundary layer correction in a reduced model.

The interior, thermal wind and Ekman layer components are respectively denoted by superscripts $(o)$, $(m,\pm)$ and $(i,\pm)$ that when combined form the composite solution,
\beginar
\label{eq:vgvEsplit}
    \boldsymbol{v} &=& \boldsymbol{v}^{(o)} \lb x,y,\Omega,t\rb + \boldsymbol{v}^{(m,+)} \lb x, y, 0, \eta^-, t \rb + \boldsymbol{v}^{(m,-)} \lb x, y, 1, \eta^+, t \rb \\
    && + \boldsymbol{v}^{(i,+)} \lb x, y, 0, \mu^-, t \rb + \boldsymbol{v}^{(i,-)} \lb x, y, 1, \mu^+, t \rb . \nonumber 
\endar
Here, $+\ (-)$ refer to the lower (upper) boundary. Thus
$\eta^+=\eps^{-1}\Omega$ and $\eta^- = \eps^{-1}(1-\Omega)$, both $\ge 0$, are the middle coordinates within the thermal wind layer which in dimensional units translates to $\mathcal{O}(Ek^{1/3}H)$ scales. Similarly, 
$\mu^+=\eps^{-3/2}\Omega$ and  $\mu^- = \eps^{-3/2}(1-\Omega)\ge 0$ are the fast coordinate within the Ekman layer which in dimensional units translates to $\mathcal{O}(Ek^{1/2}H)$ scales. The dependency on the co-latitudinal Ekman number implies that the boundary layer depths increase with $\vartheta_f$ by a factor of $(\cos(\vartheta_f))^{-1}$.

To proceed, we employ a multiple scale expansion in the axial direction
\beginar
\pd{\Omega}\mapsto \pd{\Omega}  + \delta \eps^{-1} \pd{\eta} + \delta \eps^{-3/2}  \pd{\mu}
\endar
where for convenience, we define
\begin{equation}
  \delta =   
    \begin{cases}
    +1 & \mbox{bottom layer } (\Omega = 0)\\
    -1 & \mbox{top layer } (\Omega = 1)
    \end{cases}
\end{equation}
such that the fast coordinate derivatives may be compactly interpreted.
Each region of the fluid layer may be accessed by the following actions for the outer $(o)$, middle $(m)$, and inner $(i)$ limits on \eqref{eq:vgvEsplit}:
\begin{subequations}
\beginar
\lim \lb \boldsymbol{v} \rb^{o} &=& \lim_{\mu\rightarrow\infty \atop \eta\rightarrow\infty } \lb \boldsymbol{v}\rb =  \boldsymbol{v}^{(o)} \\
&\implies& \lim \lb  \boldsymbol{v}^{(o)} \rb^{o}=  \boldsymbol{v}^{(o)},\ \  \lim \lb  \boldsymbol{v}^{(m)},   \boldsymbol{v}^{(i)} \rb^{o}=\boldsymbol{0},
\nonumber \\
\lim \lb \boldsymbol{v} \rb^{m} &=& \lim_{\mu\rightarrow \infty \atop \Omega\rightarrow 0} \lb \boldsymbol{v}\rb =  \boldsymbol{v}^{(o)}(0) +  \boldsymbol{v}^{(m)}  \\
&\implies& \lim \lb   \boldsymbol{v}^{(o)} \rb^{m}=\boldsymbol{v}^{(o)}(0),\ \  \lim \lb  \boldsymbol{v}^{(m)} \rb^{m}=\boldsymbol{v}^{(m)},
\ \  \lim \lb  \boldsymbol{v}^{(i)} \rb^{m}=\boldsymbol{0},
 \nonumber \\
 \lim \lb \boldsymbol{v} \rb^{i} &=& \lim_{\eta\rightarrow 0 \atop \Omega\rightarrow 0} \lb \boldsymbol{v}\rb =  \boldsymbol{v}^{(o)}(0) +  \boldsymbol{v}^{(m)}(0) +  \boldsymbol{v}^{(i)}   \\
&\implies& \lim \lb   \boldsymbol{v}^{(o)} +  \boldsymbol{v}^{(m)}  \rb^{i}=\boldsymbol{v}^{(o)}(0) + \boldsymbol{v}^{(m)}(0),\ \  \lim \lb  \boldsymbol{v}^{(i)} \rb^{i}=\boldsymbol{v}^{(i)}.
\nonumber
\endar
\end{subequations}
Identical expressions hold for the upper middle and inner layers located at $\Omega=1$. By definition,  middle variables are identically zero in the outer region, while  inner variables are  identically zero in both the outer and middle regions. Contributions to the  inner region from the outer and middle variables, and, the middle region from outer variables are obtained by Taylor expanding variables in the relevant boundary layer coordinate and taking its limit to zero. 
The composite variables  \eqref{eq:vgvEsplit} (i.e., the superposition of the geostrophic, thermal wind and Ekman layer components) must satisfy boundary conditions (\ref{eq:wthetaBCs}) and either (\ref{eq:noslipandstressfreeBCs}a) or (\ref{eq:noslipandstressfreeBCs}b) at leading order as $\eps\rightarrow0$. 

\subsection{Ekman Layers (inner layers)}
\label{sec:Ekman}
In order to deduce the system of equations satisfied by $\boldsymbol{v}^{(i)}$,  the inner limit of the iNSE (\ref{eq:covariantbasisequationsrescaledfull}) must be taken and the outer and middle contributions subtracted out.  Given that $\ub_\perp^{(o)}\equiv (u^{(o)},v^{(o)}) = \mathcal{O}(1)$, $\ub_\perp^{(m)} = \mathcal{O}(\eps)$  (see following subsection on the middle layer analysis),
together with boundary conditions \eqref{eq:wthetaBCs} and (\ref{eq:noslipandstressfreeBCs}),
the dominant contributions that may participate in the analyses are deduced from (\ref{eq:covariantbasisequationsrescaledfull}) as  
\begin{subequations}
\label{eq:leadingorderEkman}
\beginar
-v^{(i)}  &\approx& \pd{\mu}^2 u^{(i)},\\
\frac{1}{\eta_3^2}u^{(i)}-\frac{\gamma}{\eps^{1/2}} \delta \pd{\mu} p^{(i)} &\approx& \pd{\mu}^2 v^{(i)},\\
-\gamma u^{(i)} +\frac{1}{\eps^{1/2}} \delta\pd{\mu} p^{(i)} &\approx& \pd{\mu}^2 w^{(i)},\\
\pd{x}u^{(i)} +\pd{y} v^{(i)} +\eps^{-1/2}\delta \pd{\mu} w^{(i)} &\approx& 0.
\endar
\end{subequations}
 This follows from the observation that $(p^{(i)}, w^{(i)}) = o\lb \ub_\perp^{(i)} \rb$ within the inner layer. This holds for all non-equatorial values $\gamma = o(\eps^{-1/2})$. 

The no-slip condition, (\ref{eq:noslipandstressfreeBCs}a) and  incompressibility (\ref{eq:leadingorderEkman}d)
simply imply
\begin{subequations}
 \label{eq:uENSscaling}
\begin{equation}
   \ub_{\perp NS}^{(i)} = \mathcal{O}(1), \quad  w^{(i)}_{NS}= \mathcal{O}(\eps^{1/2}).
\end{equation}
The dominant contributions from momentum equations (\ref{eq:leadingorderEkman}b,c) then reveal
\begin{equation}
 p_{NS}^{(i)}  = \left\{
 \begin{array}{cc}
    \mathcal{O}(\eps^{1/2} \gamma) & \mbox{for}\ \gamma > \mathcal{O}(\eps^{1/2})     
    \\ \\
    \mathcal{O}(\eps)& \mbox{for}\ \gamma \leq \mathcal{O}(\eps^{1/2})  
 \end{array}
 \right . .
\end{equation}
\end{subequations}
For stress-free conditions, the dominant terms in (\ref{eq:noslipandstressfreeBCs}b) imply that we must take
\begin{subequations}
   \label{eq:uESFscaling}
\begin{equation}
\hspace{-1em} \ub_{\perp SF}^{(i)}  = \left\{
 \begin{array}{cccr}
    \mathcal{O}(\eps^{1/2}\gamma) & \mbox{for}\ \gamma > \mathcal{O}(\eps) & s.t. & -\gamma\pd{y}\ub_{\perp}^{(o)} 
    +\delta \eps^{-1/2} \pd{\mu}\ub_{\perp}^{(i)} \approx0    
    \\ \\
    \mathcal{O}(\eps^{3/2})& \mbox{for}\ \gamma = \mathcal{O}(\eps)  &  s.t. & \lb -\gamma\pd{y}
     + \eps \pd{\Omega}\rb \ub_{\perp }^{(o)} 
    +\delta \eps^{-1/2} \pd{\mu}\ub_{\perp}^{(i)} \approx0   
    \\ \\
    \mathcal{O}(\eps^{3/2})& \mbox{for}\ \gamma= o(\eps)  & s.t. & \eps \pd{\Omega} \ub_{\perp }^{(o)} 
    +\delta \eps^{-1/2} \pd{\mu}\ub_{\perp}^{(i)} \approx0 
 \end{array}
 \right . 
\end{equation}
along with the dominant contributions from momentum equations (\ref{eq:leadingorderEkman}b,c) that 
gives
\begin{equation}
\hspace{-2em}
 w_{\perp SF}^{(i)}  = \left\{
 \begin{array}{cc}
    \mathcal{O}(\eps \gamma) & \mbox{for}\ \gamma > \mathcal{O}(\eps)     
    \\ \\
    \mathcal{O}(\eps^{2})& \mbox{for}\ \gamma \leq \mathcal{O}(\eps)    
 \end{array}
 \right .,\quad 
  p_{\perp SF}^{(i)}  = \left\{
 \begin{array}{cc}
    \mathcal{O}(\eps \gamma\cdot\max{[\gamma,\eps^{1/2}]}) & \mbox{for}\ \gamma > \mathcal{O}(\eps)     
    \\ \\
    \mathcal{O}(\eps^{5/2})& \mbox{for}\ \gamma \leq \mathcal{O}(\eps)    
 \end{array}
 \right . 
\end{equation}   
\end{subequations}
Remarkably, irrespective of the case considered, elimination $p^{(i)}$ in  (\ref{eq:leadingorderEkman}) gives
\begin{subequations}
\label{eq:Ekmansystem}
\beginar
-v^{(i)}  &\approx& \pd{\mu}^2 u^{(i)},\\
u^{(i)} &\approx& \pd{\mu}^2 v^{(i)},\\
\pd{x}u^{(i)} +\pd{y} v^{(i)} +\eps^{-1/2}\delta \pd{\mu} w^{(i)} &\approx& 0
\endar
which is identical to existing theory for the classical upright Ekman layer \citep{hG69}, albeit now for the non-orthogonal axial coordinate representation. The first two equations, (\ref{eq:Ekmansystem}a) and (\ref{eq:Ekmansystem}b), combine to give 
\beginar 
\lb \pd{\mu}^4 +1\rb (u^{(i)},v^{(i)}) = 0
\endar 
which has the general solution  
\end{subequations}
\begin{subequations}
\beginar
 u^{(i)} &=& e^{-\mu/\sqrt{2}}\lb c_1(x,y) \cos\lb\frac{\mu}{\sqrt{2}}\rb +c_2(x,y) \sin\lb \frac{\mu}{\sqrt{2}}\rb  \rb,\\
  v^{(i)} &=& e^{-\mu /\sqrt{2}}\lb -c_1(x,y) \sin\lb\frac{\mu}{\sqrt{2}}\rb +c_2(x,y) \cos\lb \frac{\mu}{\sqrt{2}}\rb  \rb.
\endar
\end{subequations}
From integrating (\ref{eq:Ekmansystem}c), and enforcing $w^{(i)}\to 0$ as $\mu \to \infty$, we obtain (upon dropping the functional spatial dependencies on $c_i$'s for notational convenience)
\begin{equation}
\begin{split}
    w^{(i)} =\ &\frac{\delta\eps^{1/2}}{\sqrt{2}} e^{-\mu/\sqrt{2}}\lb \lb \pd{x}\lb c_1+c_2\rb+\pd{y}\lb c_2-c_1\rb \rb\cos\lb \frac{\mu}{\sqrt{2}}\rb \rb \\
    +\ &\frac{\delta\eps^{1/2}}{\sqrt{2}} e^{-\mu/\sqrt{2}}\lb \lb\pd{x} \lb c_2-c_1\rb -\pd{y} \lb c_1+c_2 \rb \rb\sin \lb \frac{\mu}{\sqrt{2}}\rb \rb.    
\end{split}
    \label{eq:wEgeneral}
\end{equation}
\normalsize
The coefficients $c_i$ may now be determined upon application of either no-slip  or stress-free boundary conditions, equations (\ref{eq:noslipandstressfreeBCs}a) or (\ref{eq:noslipandstressfreeBCs}b).

No-slip boundary conditions
\begin{equation}
       \ub_\perp^{(o)}\biggr\rvert_{\Omega = 0,1}+ \ub^{(i)}_{\perp NS} \biggr\rvert_{\mu = 0} = 
       \boldsymbol{0}
\end{equation}
yield the Ekman layer solutions 
\begin{subequations}
    \beginar
 u^{(i)}_{NS} &=& -e^{-\mu/\sqrt{2}}\lb u^{(o)} \cos\lb \frac{\mu}{\sqrt{2}}\rb+v^{(o)}\sin\lb \frac{\mu}{\sqrt{2}}\rb \rb,\\
v^{(i)}_{NS} &=& e^{-\mu/\sqrt{2}}\lb u^{(o)}\sin\lb \frac{\mu}{\sqrt{2}}\rb - v^{(o)}\cos\lb \frac{\mu}{\sqrt{2}}\rb\rb, \\
w^{(i)}_{NS} &=& \frac{\delta\eps^{1/2}}{\sqrt{2}}e^{-\mu/\sqrt{2}}\lb \lb \pd{y}u^{(o)} - \pd{x} v^{(o)}\rb \lb \cos\lb \frac{\mu}{\sqrt{2}}\rb +\sin\lb \frac{\mu}{\sqrt{2}}\rb \rb  \rb.
    \endar
    \label{eq:uEvEwENS}
\end{subequations}

For stress free boundaries, with the absence of a thermal wind layer at leading order,
\begin{equation}
 \haz \cdot\nabla   \ub_\perp^{(o)}\biggr\rvert_{\Omega = 0,1} 
+\delta \eps^{-1/2} \pd{\mu}\ub^{(i)}_{\perp SF} \biggr\rvert_{\mu = 0}=\boldsymbol{0}
\end{equation}
where 
$\haz \cdot\nabla=-\gamma \pd{y}+\eps\pd{\Omega}\equiv \mathcal{L_B}$.  This yields the solution 
\begin{subequations}
\label{eq:uEvEwESF}
\beginar
u^{(i)}_{SF} &=& \frac{\delta\eps^{1/2} }{\sqrt{2}} e^{-\mu/\sqrt{2}}\mathcal{L_B}\lb  \lb u^{(o)}+v^{(o)}\rb \cos\lb\frac{\mu}{\sqrt{2}}\rb -  \lb u^{(o)}-v^{(o)}\rb \sin\lb \frac{\mu}{\sqrt{2}}\rb \rb,\ \ \\\
v^{(i)}_{SF} &=& - \frac{\delta\eps^{1/2} }{\sqrt{2}} e^{-\mu/\sqrt{2}}\mathcal{L_B}\lb\lb u^{(o)} +v^{(o)}\rb \sin\lb\frac{\mu}{\sqrt{2}}\rb + \lb u^{(o)}-v^{(o)}\rb \cos\lb \frac{\mu}{\sqrt{2}}\rb \rb,\hspace{3em} \\
w^{(i)}_{SF} &=& - \eps  e^{-\mu/\sqrt{2}} \mathcal{L_B}\lb \lb \pd{y}u^{(o)}-\pd{x}v^{(o)}\rb \cos\lb \frac{\mu}{\sqrt{2}}\rb \rb .
\endar
\end{subequations}
Note, these solutions automatically capture the situations $\gamma=\mathcal{O}(\eps)$ and/or $\pd{y}=\mathcal{O}(\eps)$. The stress-free boundary conditions, now $\haz \cdot\nabla\ub^{(o)}_\perp = o(\eps)$, are automatically achieved to leading order without need of boundary layer corrections. Inspection of the iNSE \eqref{eq:covariantbasisequationsrescaledfull} at the boundaries 
reveal the geostrophic outer boundary constraint $\partial_\Omega p^{(o)}=o(1)$.

\subsection{The geostrophic interior \& parameterized pumping conditions. }
\label{sec:interiorandpumping}

Above, we have defined the Ekman layer  (inner)  variables $u^{(i)}$, $v^{(i)}$, and $w^{(i)}$, but we have yet to define the boundary criteria on  outer solution $\boldsymbol{v}^{(o)}$ for the interior of the domain.
Given the assumption of a geostrophic interior, for $u^{(o)}$ and $v^{(o)}$, we assert that a geostrophic balance holds thru to the impenetrable boundaries. That is, the dominant $\mathcal{O}(\eps^{-1})$ terms in (\ref{eq:covariantbasisequationsrescaledfull}a) and (\ref{eq:covariantbasisequationsrescaledfull}b), which we will define as $V^g$ and $U^g$, must balance, yielding
    \beginar
  \left.
\begin{array}{r}
    V^g \equiv v^{(o)}-\gamma w^{(o)} -\pd{x}p^{(o)}=0   \\ \\
  U^g \equiv  u^{(o)}+\pd{y} p^{(o)} = 0  
\end{array}
\right \}
\quad \mbox{on}\ \Omega = 0, 1,
\label{eq:UV}
 \endar
and for $\theta$, 
\begin{equation}
    \theta^{(o)} = 0\quad  \mbox{on}\ \Omega = 0, 1.
\end{equation}

The definitions for $w^{(i)}$ given by (\ref{eq:uEvEwENS}c) or (\ref{eq:uEvEwESF}c) do not satisfy impenetrability $w=0$, so the boundary condition on $w^{(o)}$ must compensate to ensure this remains so. Requiring 
\begin{equation}
    w^{(o)}\biggr\rvert_{\Omega = 0,1}+w^{(i)}\biggr\rvert_{\mu = 0} =0
\end{equation}
implies that 
\begin{subequations}
    \label{eq:pumpingconditionsw}
    \beginar
    w^{(o)}_{NS} =&  \displaystyle\frac{\delta\eps^{1/2}}{\sqrt{2}}\lb \pd{x} v^{(o)}- \pd{y}u^{(o)}\rb,  \quad &\mbox{on}\ \Omega = 0, 1,\\
    w^{(o)}_{SF} =&-\eps \haz \cdot\nabla   \lb \pd{x}v^{(o)}-\pd{y}u^{(o)} \rb, \quad &\mbox{on}\ \Omega = 0, 1.
    \endar
\end{subequations}
Equation (\ref{eq:pumpingconditionsw}a) for no-slip boundaries  is identical in form to the classical Ekman layer \cite{hG69}, extended to the upright QG-RBC by \cite{jCKMSV16}, and now to the $f$-plane. It illustrates that the presence of cyclonic (anticylonic)  vertical vorticity $\zeta^{(o)} = \pd{x}v^{(o)}-\pd{y}u^{(o)}>0$ ($\zeta <0$) at the boundaries result in fluid being pumped away from (suctioned into) the Ekman layer. 

Equation (\ref{eq:pumpingconditionsw}b) for  stress-free boundaries  establishes that the important criteria for pumping/suction at the boundaries is the normal gradient of vertical vorticity. Negative gradients of vertical vorticity result in fluid be pumped away from the lower boundary and suctioned into the upper boundary. The reverse is true for positive gradients.

\subsection{Evidence for a Thermal Wind Layer} 
\label{sec:TW}

We first recall from the discussion on Equation \eqref{eqn:Roc} that validity of the QG-RBC system requires $\Ra=o(\eps^{-1})$, $Ro=o(1)$ and $  \zeta_0^{(o)} = o(\eps^{-1})$. At  $\Omega = (0, 1)$, the parameterized Ekman velocity boundary conditions (Equation \eqref{eq:pumpingconditionsw}) imply an outer thermal response satisfying 
\begin{subequations}
\label{eq:QGRBC_T}
    \beginar
 \pd{t} \theta^{\prime(o)}_1 +J\left[\psi^{(o)}_0,\theta^{\prime(o)}_1\right] + w_0^{(o)}\left (\partial_\Omega \overline{\Theta}_0 -1\right) &=&\frac{1}{\sigma}\nabla_\perp^2 \theta^{\prime(o)}_1,
\endar
along with associated thermal variance equation
\beginar
\label{eqn:tvar}
\frac{1}{2} \pd{t} \overline{\lb\theta^{\prime(o)}_1\rb^2} + \overline{\lb w_0^{(o)}\theta^{\prime(o)}_1\rb } \left (\partial_\Omega \overline{\Theta}_0 -1\right) &=&-\frac{1}{\sigma}\overline{\vert \nabla_\perp \theta^{\prime(o)}_1\vert^2}.
\endar
\end{subequations}
From a statistically stationary viewpoint, this implies $\theta_1^{\prime(o)}=\mathcal{O}( \sigma w_0^{(o)}\partial_\Omega \overline{\Theta}_0)$ and convective flux $\overline{w_0^{(o)}\theta^{\prime(o)}_1}\sim 
\sigma w_0^{(o)2}\partial_\Omega \overline{\Theta}_0$ on $\Omega = (0, 1)$. The stationary mean temperature equation implies 
\beginar 
\sigma \overline{w_0^{(o)}\theta^{\prime(o)}_1}  -  \partial_{\Omega} \overline{\Theta}_0  = Nu -1 
    \endar
where  $Nu$ is the Nusselt number characterizing the non-dimensional heat transport. It follows that the convective flux due to Ekman pumping remains subdominant to  
heat transport by conduction, i.e., $\partial_{\Omega} \overline{\Theta}_0\sim Nu$ and $\overline{w_0^{(o)}\theta^{\prime(o)}_1}=o(Nu)$, provided
 \beginar
 \label{eqn:pbound}
\left .  w_0^{(o)} \right \vert_{\Omega=0,1} = \left \{
\begin{array}{ccc}
 \mathcal{O} ( \eps^{1/2} \zeta_0^{(o)} ) = o( 1 ) &   & \mbox{NS} \\ \\
\mathcal{O} ( \eps        \zeta_0^{(o)}   ) =o(1) &   &   \mbox{SF}
\end{array}
\right . .
\endar
If this holds, the above estimate for thermal fluctuations on the boundary implies $\theta^{\prime(o)}_1 = o(1)$. 
Hence,  thermal corrections are not required and thermal-wind boundary layers is not necessary. Within the range of validity of the QG-RBC, criterion (\ref{eqn:pbound}b) is always satisfied asymptotically on stress--free boundaries.  For no-slip boundaries the criteria is violated when
 \beginar
  \mathcal{O} ( \eps^{-1/2}) \le   \zeta_{0,NS}^{(o)} <  \mathcal{O} ( \eps^{-1})
   \quad \implies \quad 
 \mathcal{O} (1) \le  \left. \theta^{(o)}_{1,NS}  \right \vert_{\Omega=0,1}\le \mathcal{O}  (\eps^{-1/2})
 \endar
 assuming $Nu=\mathcal{O}(1)$.

 Rectifying the ability to satisfy thermal boundary conditions for no-slip boundaries thus requires the presence of a middle layer, i.e., a thermal wind boundary layer.  The middle limit of the iNSE (\ref{eq:covariantbasisequationsrescaledfull}) must be taken and the outer contribution subtracted out. This simplifies to 

\begin{subequations}
\label{eqn:twind}
\beginar
-  v_1^{(m)} + \pd{x}p_2^{\prime(m)} &=&0,\\
   u_1^{(m)} + \pd{y}p_2^{\prime(m)} &=& 0,\\
 \pd{\eta}p_2^{\prime(m)} & = &  \ \frac{\Ra }{ \sigma} \theta_1^{\prime(m)},\\
\pd{t}\ \theta_1^{\prime(m)} + \ub_{0}^{(o)} \cdot \nabla \theta_1^{\prime(m)}  
 + w_0^{(o)} \pd{\eta}  \overline{\Theta}_1^{(m)} -\overline{  w_0^{(o)} \pd{\eta} \theta_1^{\prime(m)} }
&=& \frac{1}{\sigma}\nabla^2 \theta_1^{\prime(m)},\\
\pd{x}u_1^{(m)} +\pd{y}  v_1^{(m)} &=& 0.
\endar
\end{subequations}
where $w^{\prime(m)}_1 \equiv0$. Thus rectification to support $\theta'_1=0$ on boundaries drives a thermal wind layer as identified by  (\ref{eqn:twind}a-c).

\subsection{The Composite QG-RBC}
\label{sec:CQGRBC}
Following \cite{jCKMSV16}, the system of equations for the outer and middle regions can be reconstituted to form the Composite QG-RBC (CQG-RBC) on the $f$-plane.
\begin{subequations}
\label{eq:CQGRBC}
    \beginar
\pd{t}\nabla_\perp^2\psi_0 +J\left[\psi_0,\nabla_\perp^2\psi_0\right] -\pd{\Omega}W_0 +\gamma \frac{\Ra}{\sigma} \pd{x}\theta'_1  &=& \nabla_\perp^2 \nabla_\perp^2 \psi_0,\\
 \pd{t} W_0 +J\left[\psi_0,W_0\right]  +\pd{\Omega} \psi_0 &=& \nabla_\perp^2 W_0 +\frac{\Ra}{\sigma} \theta'_1,\hspace{2em}
     \endar
     \beginar
 \pd{t} \theta'_1 +J\left[\psi_0,\theta'_1\right] + \eps \nabla_\perp\cdot \lb \ub_{1\perp} \theta'_1 \rb 
 + \underline{\eps \pd{\Omega}  \lb w_0  \theta'_1 - \overline{  w_0  \theta'_1  }\rb}
 + w_0 \left (\partial_\Omega \overline{\Theta}_0 - 1\right) = &&\hspace{4em} \nonumber \\
  \frac{1}{\sigma}\nabla^2 \theta'_1,&&\\
 \partial_\Omega \lb\overline{w_0\theta'_1} \rb = \frac{1}{\sigma} \partial_{\Omega\Omega} \overline{\Theta}_0,\hspace{4em} &&\\
 \nabla_\perp\cdot \ub_{1\perp} + \pd{\Omega} w_0 = 0 \hspace{7.7em}&&
       \endar
\end{subequations}
along with pumping boundary conditions \eqref{eq:pumpingconditionsw} and fixed temperature conditions $\overline{\Theta}_0=\theta'_1=0$. Note $\nabla^2 =\nabla^2_\perp +\eps^2 \partial_{\Omega\Omega}$.
All variables are now interpreted as composite variables, namely
\beginar
\psi^{(c)}_0 = \psi^{(o)}_0 + \eps  \psi^{(m)}_1, \ \ 
w^{(c)}_0 = w^{(o)}_0, \ \ 
\theta^{\prime(c)}_1 = \theta^{\prime(o)}_1+ \theta^{\prime(m)}_1, \ \
\overline{\Theta}^{(c)}_0 = \overline{\Theta}^{(o)}_0+ \eps\overline{\Theta}^{(m)}_1.\qquad
\endar
For convenience, the superscript $(c)$ has been dropped.

We remark that the prior sub-section has established that in the presence of stress-free boundaries, pumping conditions result in $\theta^{\prime(o)}_1=0$ on the boundaries due to the absence of a middle thermal-wind layer. This occurs because pumping velocities remain weak within the quasi-geostrophic limit. In this situation, the underlined term above is subdominant and $\nabla^2\rightarrow \nabla^2_\perp$ such that the 
CQG-RBC and QG-RBC become equivalent. This alludes to the expectation that results should be indistinguishable between the CQG-RBC model with parameterized stress-free pumping conditions and QG-RBC model with impenetrable boundaries. Indeed this finding is validated in the results section.

\section{Linear Stability}
\label{sec:linstab}
The prior section deduced the parameterized pumping boundary conditions associated with either stress-free or no-slip mechanical boundary conditions. In this section, the marginal stability problem for the onset of steady convection in the RRBC configuration is formulated using three linearized model systems: the iNSE defined in \eqref{eq:NSE} and the two asymptotically reduced models outlined in \eqref{eq:QGRBC} and \eqref{eq:CQGRBC}, respectively, the QG-RBC and CQG-RBC models. Table \ref{Table:EquationSets} summarizes these model systems along with associated physical or pumping boundary conditions.
{
\footnotesize{
\begin{table}
\begin{center}
\begin{tabular}{|l|c|c|l|}
\hline
\ \ Model \ \ & Thermal & Kinematic &\hspace{2.0em} Mechanical \\ 
\hline
 iNSE       & &  & NS:\hspace{2em} $u,v=0$ \\ &&& \\
 Eq.\eqref{eq:covariantbasisequationsrescaledfull} with & $\theta'=0$  & $w=0$   & or \\ &&& \\
physical b.c. & & & SF:\hspace{2em}  $\haz\cdot\nabla (u,v) =0$ \\
 \hline
 iNSE           & & \hspace{-1.3em}  NS:\hspace{2em}  $ w^{(o)}=  \displaystyle{\frac{\delta\eps^{1/2}}{\sqrt{2}}\lb \pd{x} v^{(o)}-\pd{y}u^{(o)}\rb}$ & \\
 Eq.\eqref{eq:covariantbasisequationsrescaledfull} with  & $\theta^{\prime(o)}=0$ & \hspace{-17.5em}  or & \hspace{1.125em}  $U^{(o)g},V^{(o)g}=0$    \\
parameterized b.c. & &  SF:\hspace{2em}  $ w^{(o)}_{SF} = -\eps \haz \cdot\nabla   \lb \pd{x}v^{(o)}-\pd{y}u^{(o)} \rb$ &\\
 \hline
QG-RBC Eq.\eqref{eq:QGRBC} & & $w^{(o)}_0=0$& \\
\hline
CQG-RBC          & &  \hspace{-1.3em}  NS:\hspace{2em}  $ w^{(o)}=  \displaystyle{\frac{\delta\eps^{1/2}}{\sqrt{2}}\lb \pd{xx} + \pd{yy}\rb \psi_0^{(o)}}$ & \\
 Eq.\eqref{eq:CQGRBC} with   & $\theta^{\prime(o)}=0$ &  \hspace{-17.5em}  or  & \\
parameterized b.c.  & &  SF:\hspace{2em}  $ w^{(o)}_{SF} = -\eps \haz \cdot\nabla \lb \pd{xx}+\pd{yy}\rb \psi_0^{(o)} $ &\\
\hline
\end{tabular}
\end{center}
\caption{\small{Summary of the various fluid equations and associated boundary conditions considered for linear stability analysis: iNSE (incompressible Navier-Stokes Equations) $8^{th}$-order in $\Omega$; QG-RBC (Quasi-Geostrophic Rayleigh-B\'enard Convection equations) $2^{nd}$-order; and CQG-RBC (Composite Quasi-Geostrophic Rayleigh-B\'enard Convection equations), $4^{th}$-order. Boundary conditions are applied at $\Omega =(0,1)$ and superscript $(o)$ denotes outer variables. In the non-orthogonal coordinate representation $\haz \cdot\nabla \equiv -\gamma \pd{y} + \eps \pd{\Omega}$. $U^{(o)g}=u^{(o)}+\pd{y} p^{(o)}$ and $V^{(o)g}=v^{(o)}-\gamma w^{(o)} -\pd{x}p^{(o)}$ are the ageostrophic variables. For the Fully nonlinear problem, mean temperature boundary condition $\overline{\Theta}=0$ on  $\Omega =(0,1)$ must be added.}}
\label{Table:EquationSets}
\end{table}
}}

We seek  solutions to the linearized version of each of the aforementioned systems about the base state $\overline{\Theta}=1-\Omega$, $\ub=\theta'=0$ by substituting the normal mode ansatz
\begin{equation}
  \boldsymbol{v} =   \hat{\boldsymbol{v}}(\Omega)\exp\left( s t + i \mathbf{k}_\perp\cdot\mathbf{x}_\perp\right)
\end{equation}
for convective rolls. Here, we define the wavenumber $\mathbf{k}_\perp = (k_x,k_y)$ by its magnitude $|\mathbf{k}_\perp|= \sqrt{k_x^2+k_y^2}\equiv k_\perp$, such that  $k_x = k_\perp\cos(\chi)$, $k_y = k_\perp\sin(\chi)$
and $\tan\lb \chi\rb = k_y/k_x$. $\chi$ defines the roll orientation with $\chi=0^\circ$ for North-South rolls and $\chi=\pi/2$ for East-West rolls. Steady convective onset occurs when growth rate $s=0$ which is known to be independent of $\sigma$ \citep{sC61}. For a specified co-latitude $\vartheta_f$, we find \textit{a posteriori} that the stability domain is bracketed by north-south convective roll orientations (the gravest mode) and east-west roll orientations  (the least excitable mode). Given the uncovering of parameterized boundaries conditions, critical questions to be addressed are (i) to what extent do solutions obtained to the iNSE  with these boundary conditions agree quantitatively with those obtained when the true physical unapproximated boundary conditions are employed, (ii) how robust is this agreement across a range of finite values of  $\eps$, i.e., is Ekman pumping captured through the parameterized boundaries conditions solely responsible for the departure from the asymptotic solution obtained as $\eps \to 0$ by the QG-RBC, and separately, and (iii) what is the fidelity of the CQG-RBC that amends the QG-RBC with parameterized boundaries conditions, i.e. again, how robust is the agreement with the iNSE for finite $\eps$.

\subsection{Linear stability of the QG-RBC}
\label{sec:asymptoticlinearstab}
Fortuitously, analytic progress can be made for the linear stability problem associated with the QG-RBC.
Here, the normal mode perturbations take  the  specific form 
\begin{equation}
    \label{eq:reducedeqnsansatz}
    \begin{split}
        \theta_1 &= \hat{\theta}\sin(n\pi\Omega)h(x,y) e^{st}  +c.c.,\\
        w_0 &= \hat{w}\sin(n\pi\Omega) h(x,y)e^{st}  +c.c.,\\
        \psi_0 &= \lb \hat{\psi} \cos(n\pi\Omega) h(x,y) + \gamma \frac{1}{k_\perp^2}  \hat{w}\sin(n\pi\Omega) \pd{x} h (x,y) \rb e^{st}  + c.c.,
    \end{split}
\end{equation} 
where $h(x,y) = \exp\left(ik_x x+ik_y y\right)$. For $n = 1,2, 3, ...$, this ansatz automatically satisfies the fixed-temperature impenetrable boundary conditions given in (\ref{eq:wthetaBCs}). The appearance of amplitude $\hat{w}$ (equivalently, the component $\pd{x} w_0$)  in the ansatz for $\psi_0$ in (\ref{eq:reducedeqnsansatz}c) is evidence of non-axial buoyancy driving on the $f$-plane giving rise to a buoyancy torque that generates axial vorticity when $\gamma\ne0$.

Substitution of \eqref{eq:reducedeqnsansatz} into the linearized QG-RBC system (\ref{eq:QGRBC}) results in an eigenproblem yielding analytic expressions for the critical Rayleigh number, critical wavenumber, and maximum growth rate. 
For the case $\sigma= 1$, the characteristic polynomial for the growth rate is given by 
\begin{equation}
    \left(k_\nabla^2+s\right)  \left(k_\nabla^2 s^2+2 k_\nabla^4 s+k_\nabla^6 +\pi ^2 n^2 -\Ra\ k_\perp^2\right) =0,
\end{equation}
where 
$$k^2_\nabla\equiv|\mathbf{k}_{\nabla}|^2 = k_x^2+k_y^2/\eta_3^2 = k^2_\perp \left(1+\gamma^2\sin^2(\chi)\right)$$ 
is  the  coefficient arising from applying the Laplacian operator $\nabla_\perp^2$. The solutions are given by eigenvalues 
\begin{subequations}
\beginar
    s &=& -k^2_\nabla, \quad \\
        s&=&-k^2_\nabla\pm \frac{1}{k_{\nabla}}\sqrt{\Ra\ k^2_\perp -n^2 \pi^2}.
        \endar
        \label{eq:eigenvalues}
\end{subequations}
The first, equation (\ref{eq:eigenvalues}a), poses no stability constraint, but the second, (\ref{eq:eigenvalues}b), yields an instability for the onset of steady convection when $\Ra >\Ra_s$, where
\begin{equation}
     \Ra_s = \frac{k_\nabla^6 + n^2\pi^2}{k_\perp^2}.
    \label{eq:marginalstabasymptotic}
\end{equation}
The eigenvector containing the relative amplitudes for the linear roll solutions are given by
\begin{equation}
(\hat w, \hat \psi, \hat \theta)^T = \lb  1, -\frac{n \pi}{ k^2_\perp k_\nabla^2},
\frac{\sigma}{k_\nabla^2} \rb^T \hat w.
\label{eq:eigenvector}
\end{equation}
The smallest value on the marginal stability curve  $\Ra_s$ is the critical point
\begin{equation}
    \Ra_c = \frac{3}{2}\lb 2\pi^4\rb ^{1/3}\left(1+\gamma^2\sin^2(\chi)\right), \ \ 
   k_{\perp c} = \frac{\pi^{1/3}}{2^{1/6}\left(1+\gamma^2\sin^2(\chi)\right)^{1/2}},
    \label{eq:asymptoticRacrit}
\end{equation}
occurring when 
$n=1$. The maximum growth rate achieved by (\ref{eq:eigenvalues}b) for mode $n=1$ is 
\begin{equation}
    s_{max} = k_\nabla^2 \left( \frac{\pi^2}{2 k_\nabla^6}-1\right),
\end{equation}
and it occurs in the $\lb k_\perp,\Ra \rb$ plane along the curve 
\begin{equation}
    \Ra = \frac{\pi^2}{k_\perp^2}\lb \frac{\pi^2}{4 k_\nabla^6} +1\rb, \quad \mbox{for}\  k_\perp\leq k_{\perp c}.
    \label{eq:Ramaxgrowth}
\end{equation}
\begin{figure}
    \centering
    \includegraphics[width = .9
\linewidth]{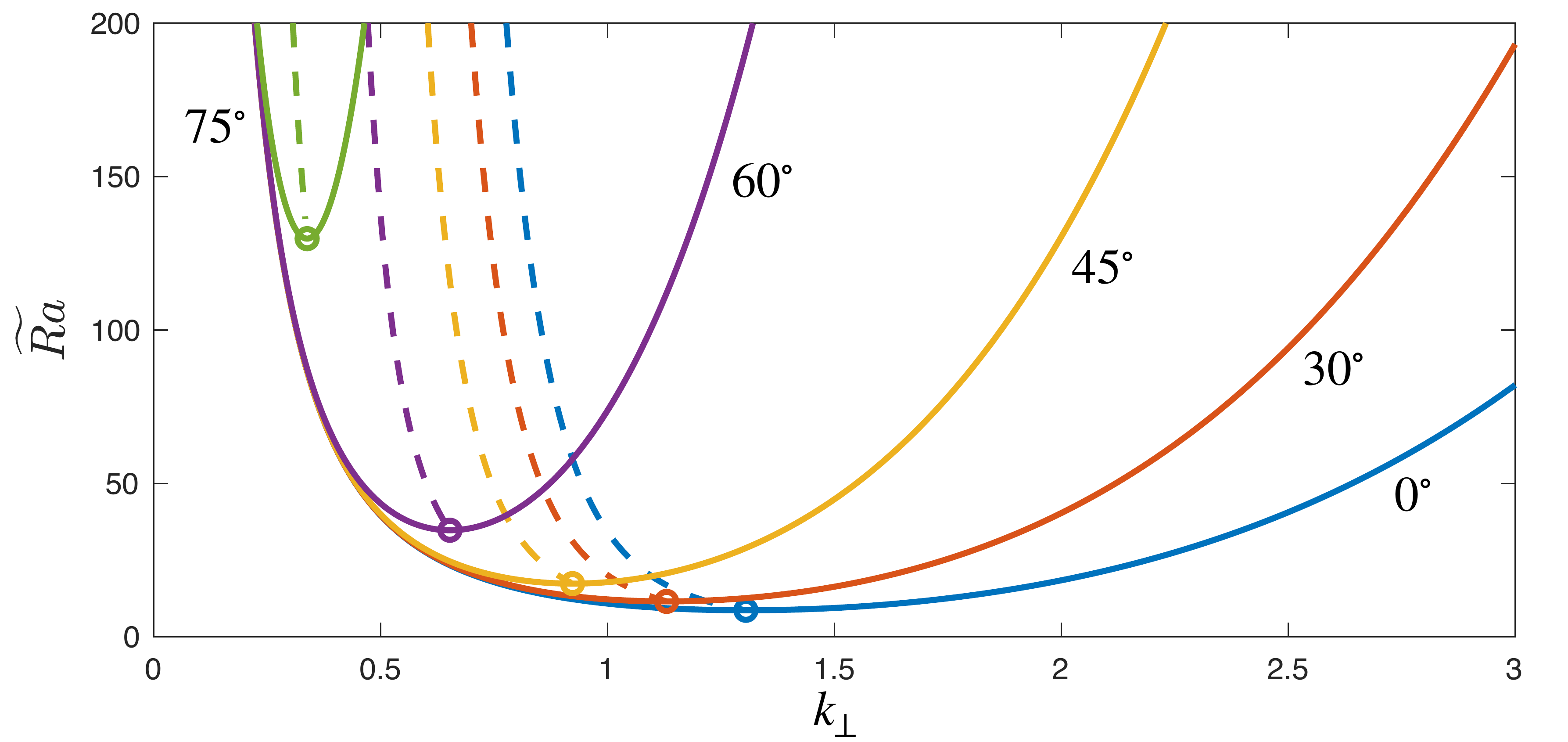}
    \caption{\small{
    Quasi-geostrophic rotating Rayleigh-B\'enard convection (QG-RBC) marginal stability curves, loci of the maximum growth rates, and critical Rayleigh and wave numbers in the $( k_\perp,\Ra )$ plane for East-West convection rolls ($\chi =\pi/2$) at various co-latitudes $\vartheta_f$ (annotated). The solid lines are the marginal stability curves defined by (\ref{eq:marginalstabasymptotic}); dashed curves are the locations of the max growth rates defined by (\ref{eq:Ramaxgrowth}); and the circles mark the critical values $(k_c,\Ra_c)$ given by (\ref{eq:asymptoticRacrit}). 
    North-South rolls with $\chi=0$ are coincident with solid blue line for all $\vartheta_f$.}}
    \label{fig:marginalstab_asymptotic}
\end{figure}
The values given by (\ref{eq:marginalstabasymptotic}), (\ref{eq:asymptoticRacrit}), and (\ref{eq:Ramaxgrowth}) in the $( k_\perp,\Ra )$ plane are plotted in figure \ref{fig:marginalstab_asymptotic} for various tilt angles $\vartheta_f$ (dashed lines). Note that for the upright case ($\gamma =0$), the expressions for the various for marginal stability properties simplify significantly, and there is no longer dependence on roll orientation $\chi$ given  $|\mathbf{k}_{\nabla}|^2 \equiv |\mathbf{k}_{\perp}|^2$. Thus the marginal stability and maximal growth rate are identical for all roll orientations, North-South through East-West rolls. These upright expressions are also identical to the North-South case $\chi=0$ for arbitrary co-latitudes $\gamma\ne0$. Thus as postulated North-South rolls provide the gravest (most unstable) mode (see blue curves plotted in Figure \ref{fig:marginalstab_asymptotic}).
East-West rolls, case $\chi = \pi/2$ are plotted in figure \ref{fig:marginalstab_asymptotic} at various $\gamma$ since they provide the bookend as the least grave or least supercritical mode.

\subsection{Departure from the Linear QG-RBC due to Ekman pumping}
\label{sec:linstabdep}
Equation \eqref{eqn:pbound} establishes the criteria for which Ekman pumping remains subdominant and the asymptotic rotating convection problem remains adequately described by the QG-RBC model with impenetrable boundaries. Recall, the reduction in the axial spatial order indicates that no mechanical boundary conditions need be prescribed. Their inclusion would require Ekman boundary layer corrections which remain passive in that they do not alter the marginal stability threshold or global heat and momentum transport properties. We have established this to be the case solely for stress-free boundary conditions. 

Given the analytic results of the prior section for the linear QG-RBC model, it is possible to estimate for no-slip boundaries when Ekman pumping becomes dominant along the marginal stability curves defined in \eqref{eq:marginalstabasymptotic} and displayed in figure \ref{fig:marginalstab_asymptotic}. This occurs when pumping velocities become $\mathcal{O}(1)$, i.e.,  $\hat w (0) = \hat w (1) = \mathcal{O}(1)$. From equations (\ref{eqn:pbound}b),  (\ref{eq:reducedeqnsansatz}c) and (\ref{eq:eigenvector}) this implies 
\beginar
\hat w = - \delta \frac{\varepsilon^{1/2}}{\sqrt{2}} k_\perp^2 \hat \psi\gtrsim\mathcal{O}(1), \qquad\mbox{s.t}\qquad
 \frac{\varepsilon^{1/2}}{\sqrt{2}}\frac{n \pi}{ k_\nabla^2}\gtrsim 1. \hspace{2em}
\endar 
Within the asymptotic validity of the QG-RBC, i.e., $\Ra=o(\varepsilon^{-1})$,  this is captured by the low wavenumber bound and transitional Rayleigh number estimates
\begin{equation}
\label{eqn:EKtrans}
 k_\perp \lesssim \varepsilon^{1/4} \lb\frac{n\pi}{\sqrt{2}}\frac{1}{\left(1+\gamma^2\sin^2(\chi)\right)}\rb^{1/2}, \qquad 
  \Ra_t \sim \eps^{-1/2}\sqrt{2}n\pi \left(1+\gamma^2\sin^2(\chi)\right). 
\end{equation}
This transition always occurs within the quasi-geostrophic regime given $\Ra_t=o(\varepsilon^{-1})$. Moreover, the transition is delayed in $\Ra_t$ and scale 
$k^{-1}_\perp$ as tilt $\gamma$ and roll orientation $\chi$ increase.

\subsection{Results: Linear Stability across Models}
\label{sec:results} 
In this section, we analyze the linear stability problem for the onset of steady convection in the RRBC.
Comparisons are made between results obtained from the iNSE and the reduced QG-RBC and CQG-RBC models solved with the various boundary condition configurations outlined in Table \ref{Table:EquationSets}. With $k_\perp$, $\chi$, $\vartheta_f$, $\eps$, and for convenience $\sigma = 1$, as input parameters, the resulting generalized eigenproblem  
is discretized with a spectral Galerkin basis constructed from Chebyshev polynomials \citep[][see Appendix~\ref{sec:NumA}]{kJ09, Ded2020}, and solved using MATLAB's sparse eigensolver package. 
We note the iNSE is solved using a vortical formulation that utilizes the geostrophic variables $U^g$ and $V^g$ \eqref{eq:UV}, thus permitting the continuance of the geostrophic constraint within the interior to the boundaries where parameterized conditions can be imposed (Details are relegated to Appendix~\ref{sec:mvort}).
\begin{figure}
    \centering
 \includegraphics[width = \linewidth]{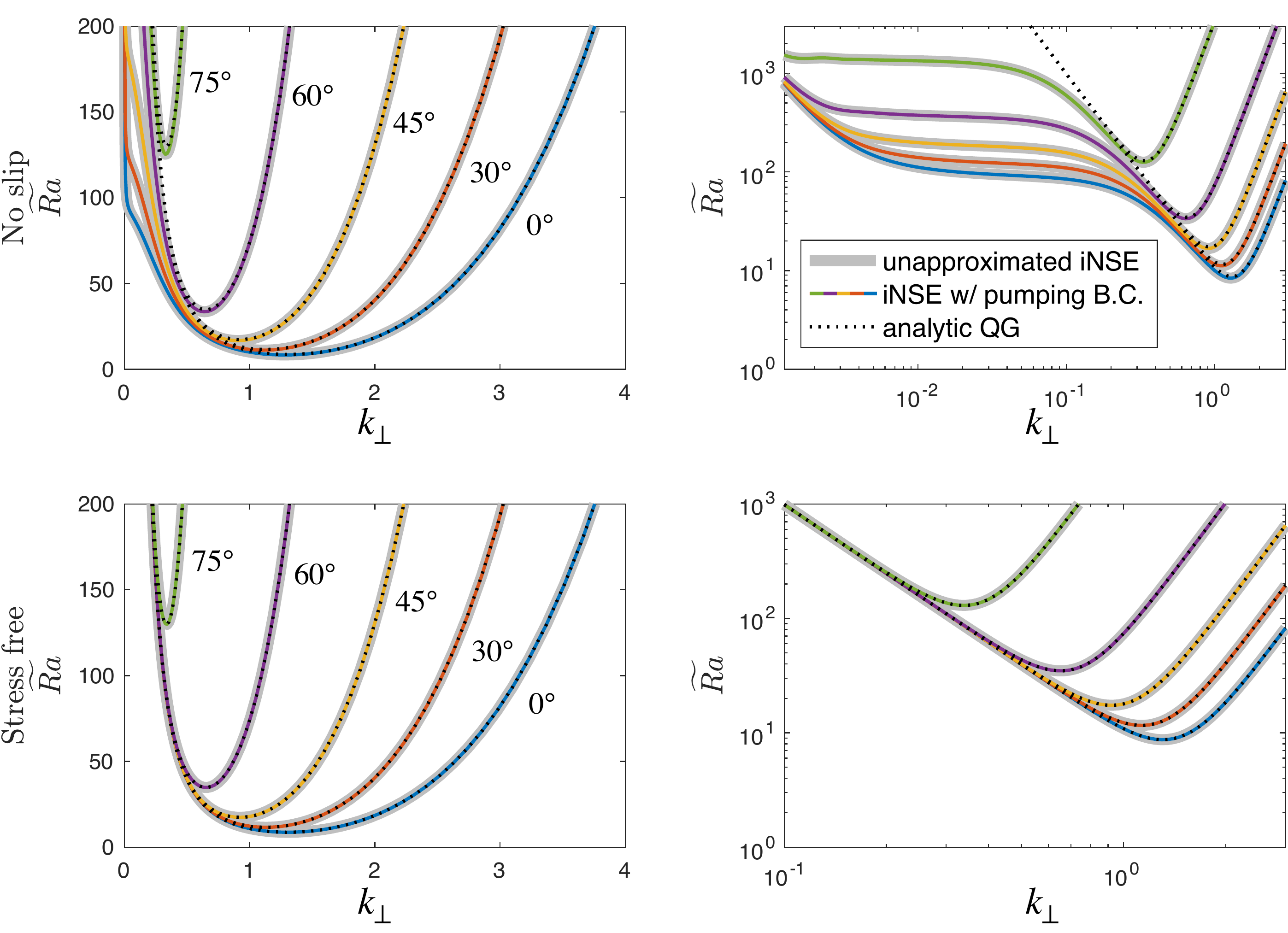}
     \caption{\small{Comparison of marginal stability curves and critical Rayleigh numbers vs wavenumber for the iNSE problem with physical (underlying translucent gray curves) and parameterized pumping boundary conditions (solid colored curves), and the analytic results from the QG-RBC problem (dotted curves) also illustrated in Figure~\ref{fig:marginalstab_asymptotic}. The case illustrated  is $\eps = 10^{-3}$ $(\ek=10^{-9})$ and $\chi = \pi/2$ (East-West rolls) at various co-latitudes $\vartheta_f$ (annotated). Plots (a) and (b) are for no-slip  boundaries. Excellent quantitative agreement exist between the iNSE and the CQG-RBC models.
    The significant impact of Ekman pumping on the onset of convection at low wavenumbers with respect to the QG-RBC model ar the result of $\mathcal{O}(\ek^{1/2})$ boundary layers is evident. 
    Plots (c) and (d) illustrate results for stress-free illustrating excellent quanititative agreement between all models.}}
        \label{fig:marginalstability_NSandSFU}
\end{figure}

\begin{figure}
\centering
   \includegraphics[width=\linewidth]{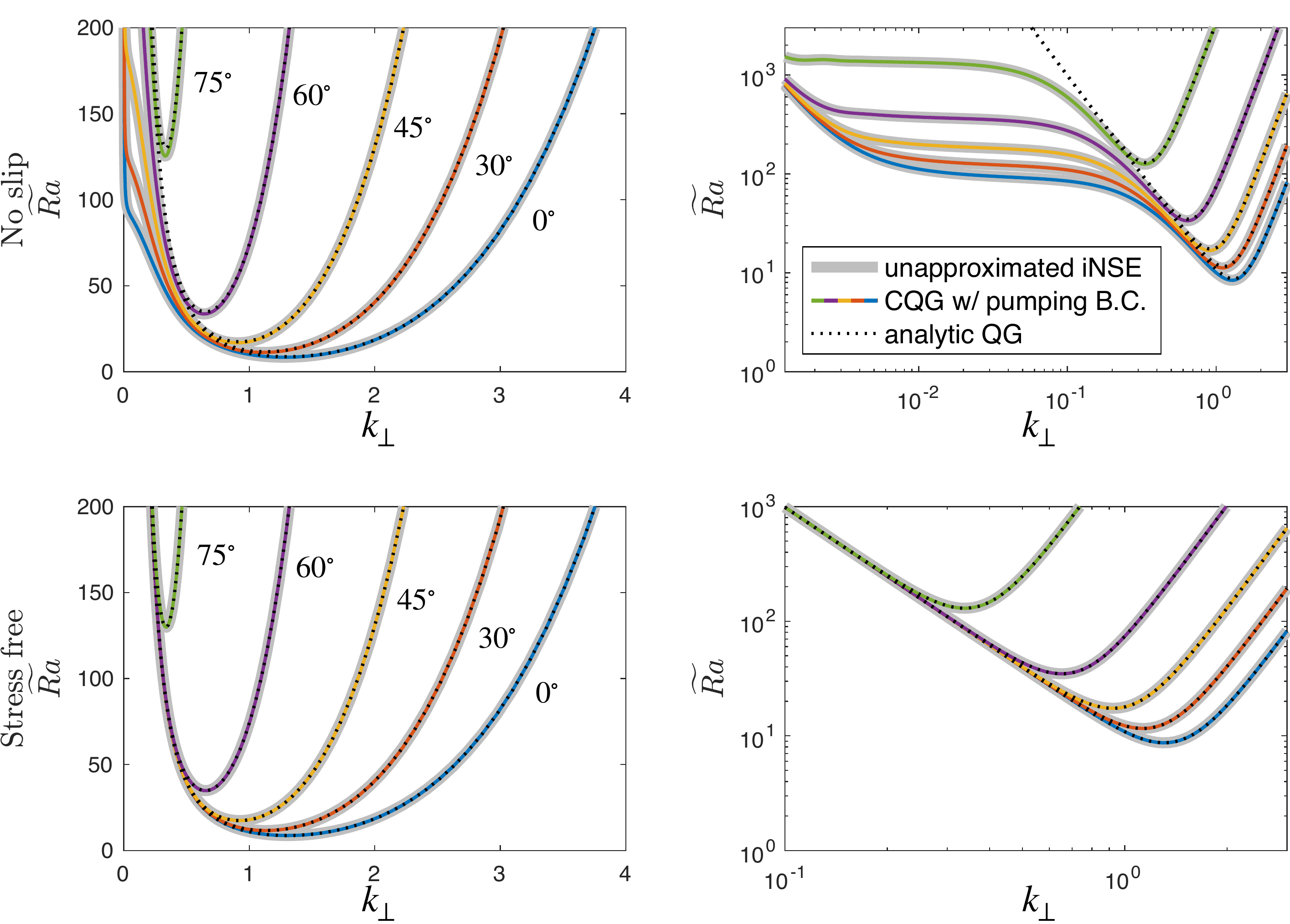}
\caption{\small{Marginal stability curves for CQG-RBC problem with parameterized pumping boundary conditions (solid colored curves), with comparison to the iNSE with physical boundary conditions  (grey translucent) and the analytic results from the QG-RBC problem (dashed curves). Additional details are as in Figure~\ref{fig:marginalstability_NSandSFU}.}}
\label{fig:CQG}
\end{figure}

Shown in figure \ref{fig:marginalstability_NSandSFU} are the  marginal stability curves computed numerically from the \textit{unapproximated} iNSE with the mechanical boundary conditions and from the iNSE  with parameterized pumping boundary conditions for East-West rolls across a range of co-latitudes. Respectively, these two model results are depicted by the \textit{underlying translucent} grey curves and solid colored curves. The representative case $\varepsilon = 10^{-3}$ $(Ek=10^{-9})$ is considered. Plots (a) and (b) illustrate the case for no-slip boundaries and plots (c) and (d) illustrate the stress-free case. Also included for reference are the asymptotic marginal stability curves obtained from the QG-RBC system  (dotted curves). 

For stress-free boundaries, we observe excellent quantitative agreement for all wavenumbers between both iNSE models and the  the asymptotic results (dotted curves) from the QG-RBC system. This is consistent with the boundary layer analysis of section~\ref{sec:boundary} showing that the $\mathcal{O}(\eps)$ pumping velocities emanating from Ekman layers adjacent to stress-free boundaries are too weak to induce corrections that alter the asymptotic predictions of the QG-RBC model. In effect, Ekman boundary layers, while necessary for the maintenance of  stress-free boundaries, remain passive. 

For no-slip boundaries results indicate excellent quantitative agreement between the two iNSE models for all wavenumbers illustrating the accuracy and fidelity of the parameterized pumping condition. However, figure~\ref{fig:marginalstability_NSandSFU} also reveals significant departures of the iNSE models from the stress-free results for low wavenumbers. Specifically, it is observed in the presence of no-slip boundaries, $\mathcal{O}(\eps^{1/2})$ pumping velocities from the Ekman layer act to further destabilize low wavenumber (large-scale) modes and thereby extends the wavenumber range for steady convective onset at a fixed $\Ra$. 
The impact of Ekman pumping on the marginal curves is more clearly illuminated in the log-log plot (b) where departures from the stress-free marginal curves first occur through an intermediate region where the $\Ra$ remains approximately constant followed by a monotonic increase in $\Ra$ with decreasing $k_\perp$ that appears to parallel the asymptotic curve that scales with $\Ra\sim k_\perp^{-2}$. Note, as predicted by equation \eqref{eqn:EKtrans}, the departure $(\Ra_t, k_{\perp t})$ from the stress-free marginal curves are respectively increasing and decreasing functions of $\vartheta_f$.

Figure~\ref{fig:CQG} illustrates that identical deductions hold for the CQG-RBC model with  parameterized pumping boundary conditions. Indeed, this asymptotic model is in excellent quantitative agreement with the both iNSE models illustrated in figure~\ref{fig:marginalstability_NSandSFU}. For stress-free boundary conditions, this result establishes the predicted equivalence between the CQG-RBC and QG-RBC models.

\begin{figure}
    \centering
\includegraphics[width = \linewidth]{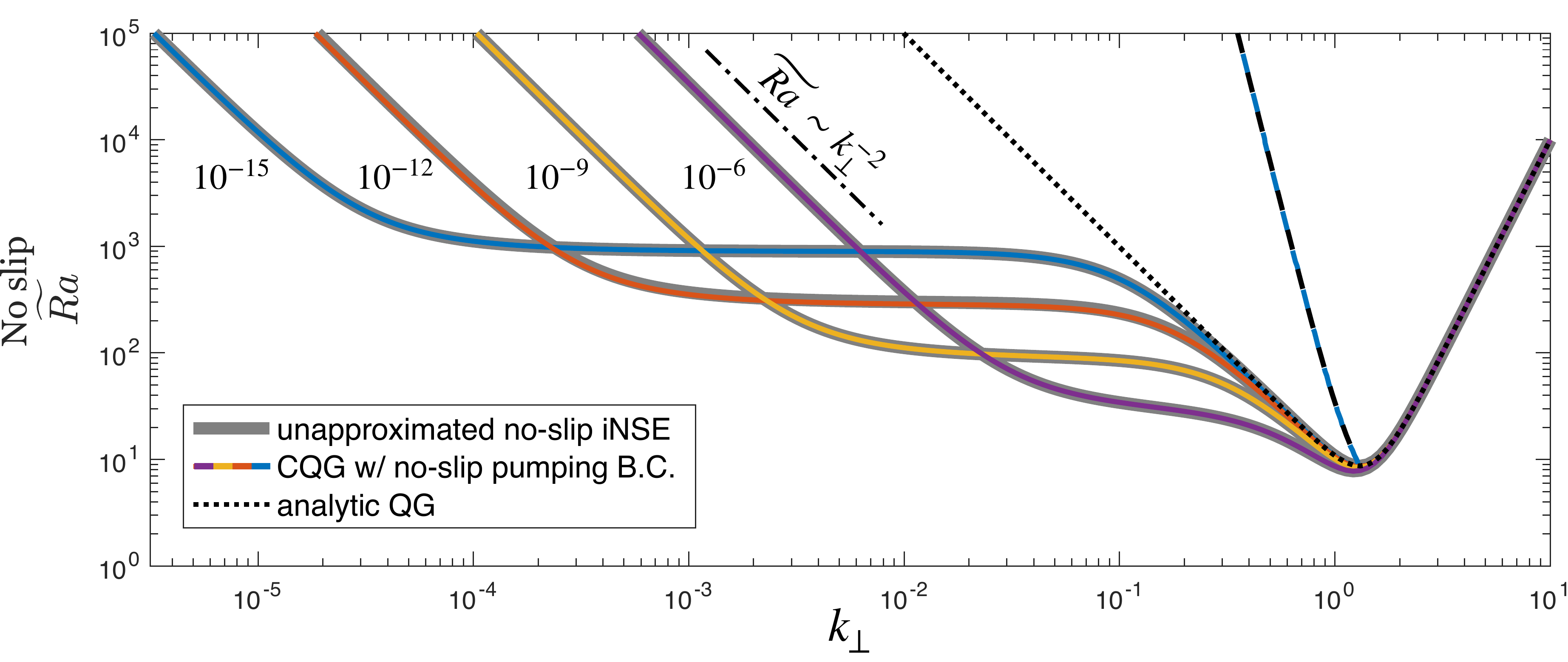}
    \caption{\small{Marginal stability curves with parameterized no-slip pumping conditions on CQG-RBC, for North-South convection rolls ($\chi=0^\circ$) and varying rotational constraint $\eps=\{10^{-2},10^{-3},10^{-4},10^{-5}\}$ (or equivalently, $\ek=\{10^{-6},10^{-9},10^{-12},10^{-15}\}$). Underlying translucent grey curves and solid colored curves represent, respectively, the iNSE and CQG-RBC models. The black dashed line follows the maximal growth rate from the analytic QG model, and the underlying blue dashed line is the maximal growth rate for iNSE with parameterized pumping at $\varepsilon = 10^{-5}$. }}
    \label{fig:NSeps}
\end{figure}
\begin{figure}
    \centering
    \includegraphics[width=\linewidth]{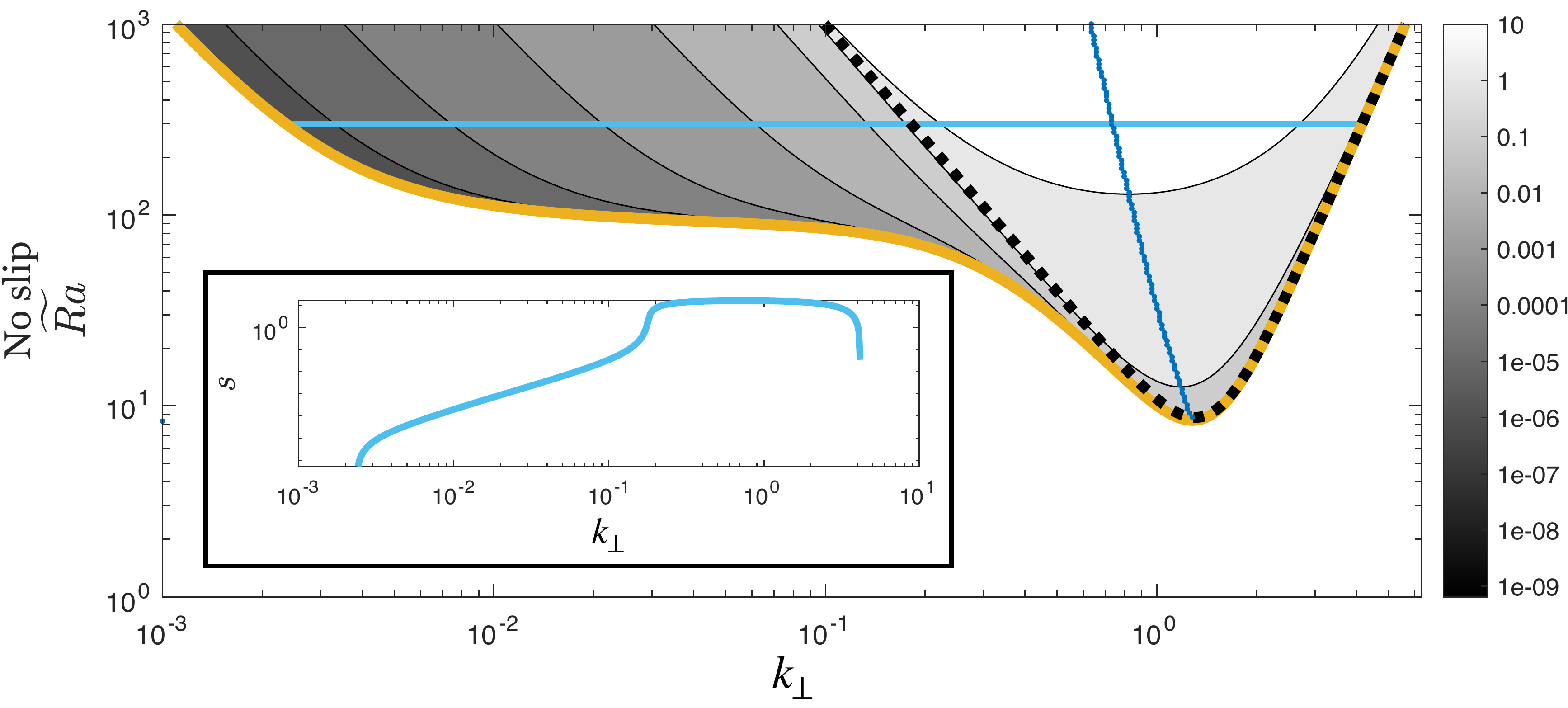}
    \caption{Contours of the growth rate for iNSE with no-slip pumping, at $\vartheta_f = 0^\circ$, $\chi = 0^\circ$, $\varepsilon = 10^{-3}$. The orange curve shows the marginal stability (where $s=0$), and the black dashed is the analytic QG marginal curve given by (\ref{eq:marginalstabasymptotic}). The blue dots mark the loci of the maximum growth rate for each value of $\widetilde{Ra}$. The light blue line shows a slice of the growth rate $s$ at $\Ra = 300$.  } 
    \label{fig:enter-label}
\end{figure}

Figure~\ref{fig:NSeps} illustrates how the marginal stability boundaries for North-South rolls, the gravest mode, change as a function of $\eps$ for the CQG-RBC and the iNSE models with parameterized pumping boundary conditions (respectfully, solid colored curves and underlying translucent grey curves). Similar results hold for differing roll orientations. It is observed that both models are in excellent quantitative agreement, as $\eps$ decreases the transition region is delayed but also extended in logarithmic range. Moreover, and quite remarkably, significant departures remain for geo- and astro-physically relevant values such as $\eps=10^{-5}$ (i.e.\ $\ek=10^{-15}$) when compared to the asymptotic QG-RBC model (dotted line). The low wavenumber departure from the QG-RBC model is consistent with the prediction detailed in equation \eqref{eqn:EKtrans} indicating transitional wavenumber $|\mathbf{k}_{\perp t}|\sim \eps^{1/4}$ and Rayleigh number $\Ra_t\sim\eps^{-1/2}$ and always occurs within the rotationally constrained regime where $\Ra=o(\varepsilon^{-1})$. Also consistent with \eqref{eqn:EKtrans} is the delay in the transitional values as a function of roll orientation $\chi$, i.e., from N-S to E-W (as seen in   figures~\ref{fig:marginalstability_NSandSFU} and \ref{fig:CQG}). 
\begin{figure}
    \centering
    \includegraphics[width = \linewidth]{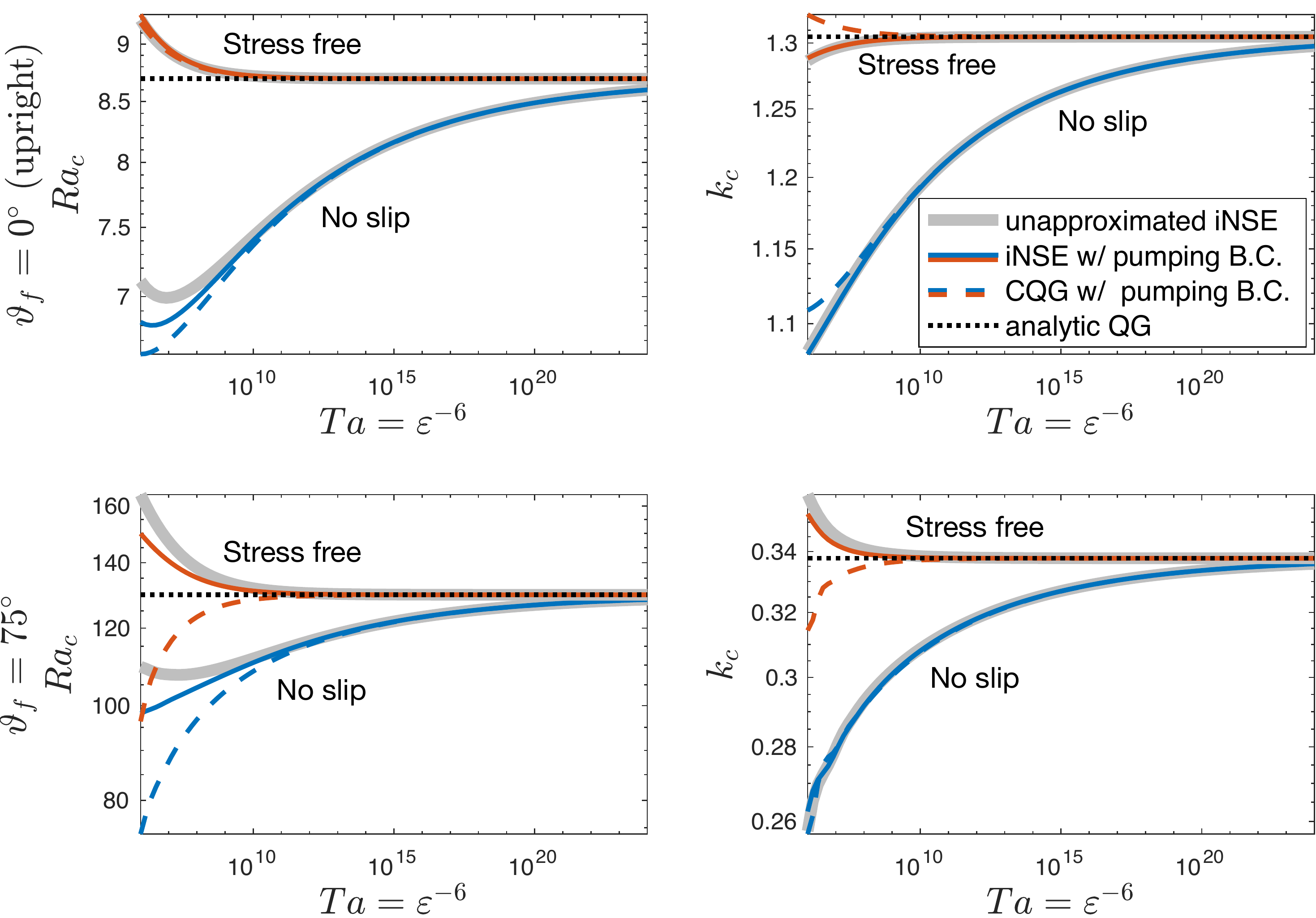}
    \caption{\small{Critical reduced Rayleigh number and wavenumber versus Taylor number $Ta$ ($=\ek^{-2} = \eps^{-6}$) for rolls ($\chi=\pi/2$) at co-latitudes $\vartheta_f = 0^\circ$ (top two panels) and East-West rolls at $\vartheta_f = 5\pi/12$ or $75^\circ$ (lower two panels). The blue curves correspond to no-slip boundary conditions and the red correspond to stress-free. Solid gray lines indicate solutions to the iNSE without approximation, and solid and dashed colored lines correspond to  solutions to the iNSE and CQG-RBC models, respectively, with pumping boundary conditions described in section~\ref{sec:interiorandpumping}. The asymptotic values for the QG-RBC given by (\ref{eq:asymptoticRacrit}) and $k_c$ are shown by the horizontal dotted line in black.}}
    \label{fig:RakvsT}
\end{figure}

Figure~\ref{fig:NSeps} also illustrates that the loci of maximal growth rate with $\Ra\propto k^{-1/8}_\perp$ remains insensitive to Ekman pumping (see dashed lines). Figure 6 expands on this point by illustrating a contour map for the growth rate in the $\Ra$-$k_\perp$ plane. One can observe that the marginal stability boundary for the asymptotic QG-RBC model strongly constrains the contours within it borders, however, the $\mathcal{O}(1)$ effect of Ekman pumping distorts the exterior contours located at  low wavenumbers. The inset illustrates a cross-section of the growth rate at fixed $\Ra=300$.

Figure \ref{fig:RakvsT} illustrates the asymptotic robustness of the parameterized boundary conditions by  tracking the minimum critical values $(\Ra_c,k_c)$ as a function of $\eps$ (specifically the Taylor number $Ta=\ek^{-2}=\eps^{-6}$) for the sample colatitude $\vartheta = 75^\circ$. It can seen that parameterizing the Ekman layer with pumping boundary conditions \eqref{eq:pumpingconditionsw} quantitatively captures the departure from the asymptotic QG-RBC value (horizontal dashed line) for the onset of convection to relatively large $\eps$ (i.e. small $Ta$) for all models. In the pertinent limit $\eps \to 0$, the critical values approach the asymptotic result albeit slowly in the case of no-slip boundaries. 
For all boundaries, one may visually observe discernible differences between the iNSE with unapproximated boundary conditions  and the iNSE with pumping boundary conditions around $Ta=10^{10}$ (i.e., $\eps\sim 10^{-5/3}, \eps\sim 10^{-5}$). 
We also observe that results from the asymptotic CQG-RBC model is in excellent quantitative agreement with those obtained from the iNSE. However, as $\eps$ becomes large, departure from the iNSE model with exact boundary conditions occurs in an opposite manner to its iNSE counterpart with pumping boundary conditions. This may attributed to the absence of vertical momentum diffusion and the unbreakable constraint of geostrophy in the CQG-RBC model.

\begin{figure}
    \centering
    \includegraphics[width = \linewidth]{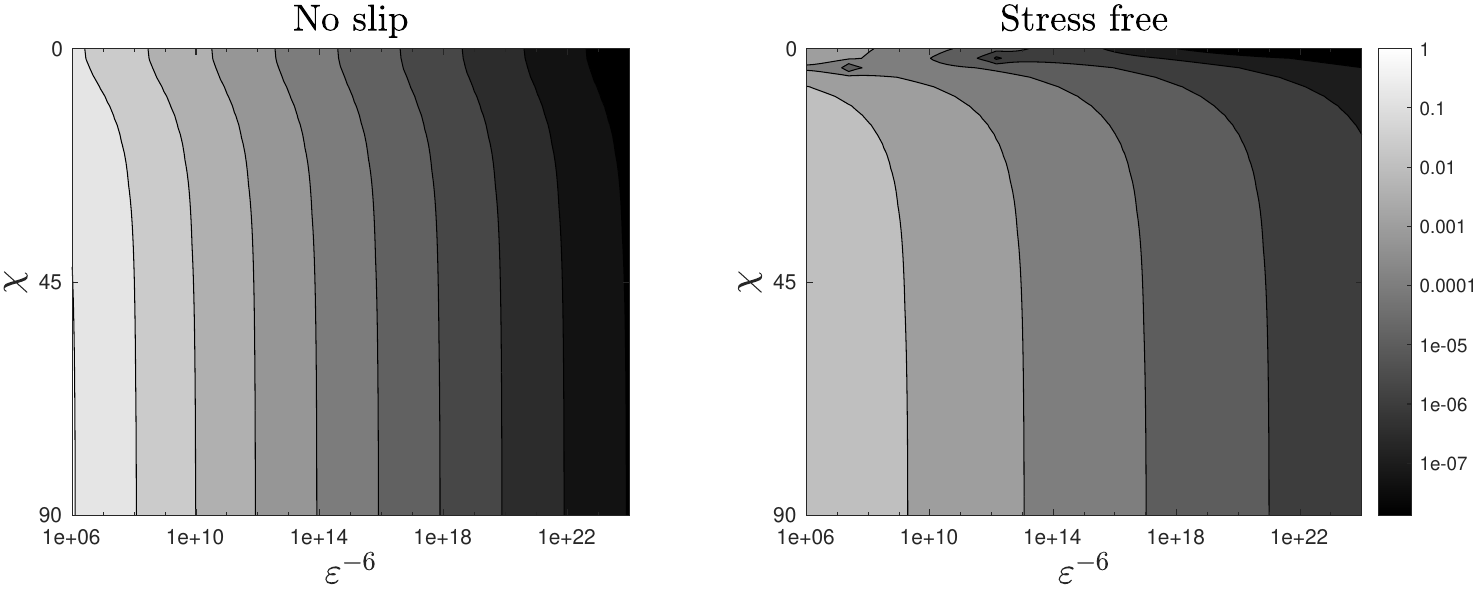}
    \caption{\small{Relative error in critical Rayleigh number between the unapproximated iNSE and the iNSE with parameterized pumping, for $\vartheta_f = 5\pi/12$ (or $75^\circ$), plotted over a range of wavenumber angles $\chi$ and Taylor numbers $\eps^{-6}$. No-slip boundary conditions (left) and stress-free conditions (right).}}
    \label{fig:relativeerror}
\end{figure}

A broader measure of the relative error between the critical onset of convection for the unapproximated iNSE problem and that with a parameterized pumping as function  of roll orientation $\chi$ and $\eps$ is shown in figure \ref{fig:relativeerror} for co-latitude $\vartheta_f = 75^\circ$. We observe that the error decays with $\eps$ across all roll orientations $\chi$. For no-slip boundaries, (plot (a)), we observe that the relative error is insensitive as a function of $\chi$ with an evolution to slightly greater accuracy occurring in the vicinity of north-south rolls $\chi< 15^\circ$. This is even more pronounced in the stress-free (plot(b)) case, but we observe a certain degree of non-monotonicity  near the top of the plot at $\chi = 0$.

\begin{figure}
    \centering
    \includegraphics[width = \linewidth]{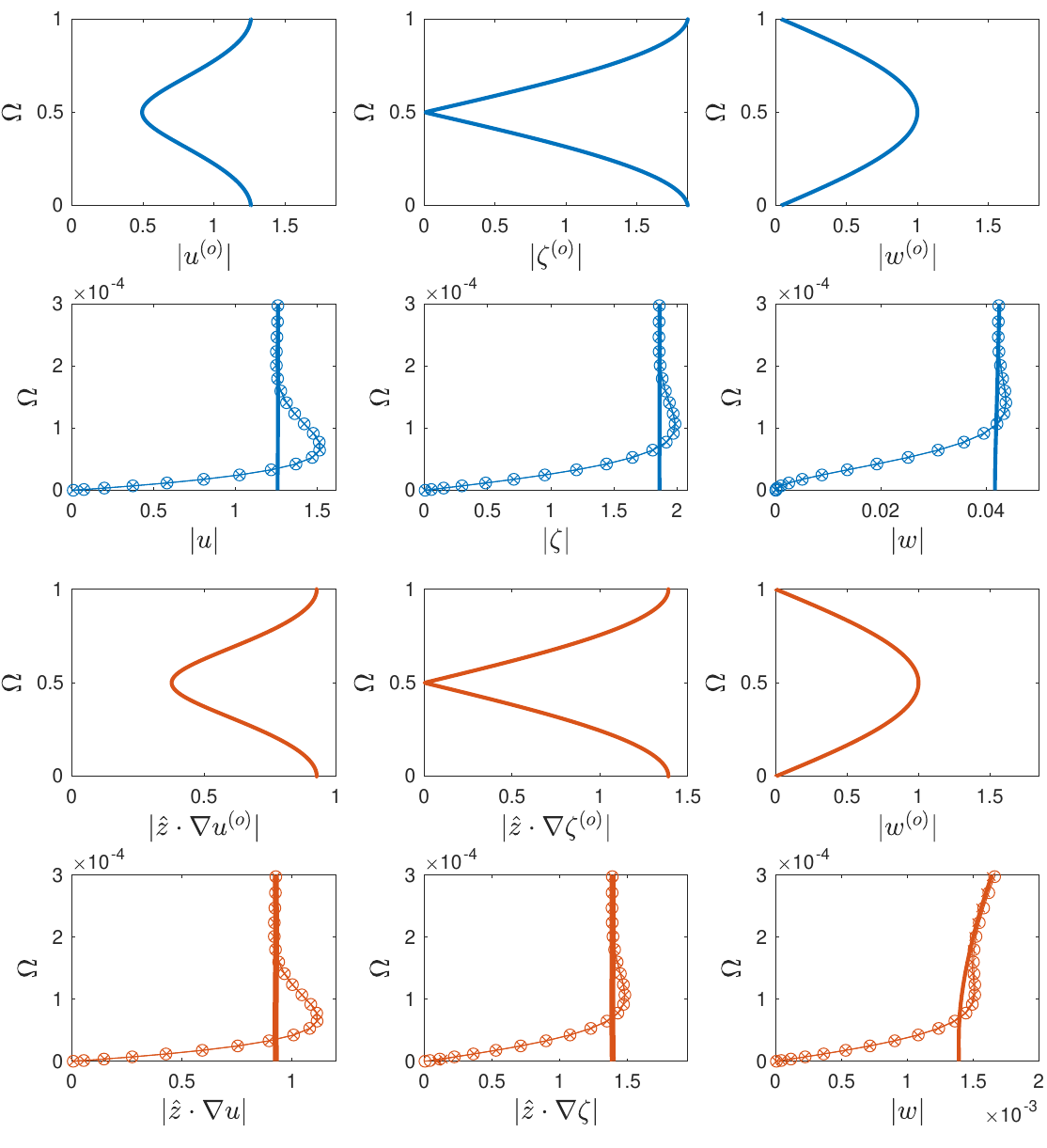}
    \caption{\small{
    Profiles at the critical Rayleigh and wavenumbers for $\vartheta_f = \pi/4$, $\chi = \pi/4$, and $\eps = 10^{-3}$. Upper two rows (blue curves) correspond to no-slip boundary conditions, and lower two rows (red curves) correspond to stress-free boundary conditions. The solid lines show the interior solution on the full domain, and the second and fourth rows show the Ekman boundary layer varying on the $\mathcal{O}(\eps^{3/2})$ scale. The open circles are the numerically computed full problem, the solid line is the numerically computed interior solution with pumping boundary conditions, and the $\times$'s are the composite solution (the numerically computed interior plus the analytic boundary layer).}} 
    \label{fig:criticalprofiles}
\end{figure}

The eigenfunctions at the critical Rayleigh and wavenumber are shown for a mid-latitude, $\vartheta_f = \pi/4$, $\chi=\pi/4$, and $\eps = 10^{-3}$ in figure \ref{fig:criticalprofiles}. 
This figure represents a direct illustration of the relaxation of spatial resolution constraints as a result of the utilization of pumping boundary conditions.
Only the outer (i.e., interior) solution, $\boldsymbol{v}^{(o)}$ is plotted on the full domain (first and third rows) since the full iNSE problem and interior iNSE problem with pumping boundary conditions are visually indistinguishable at this value of $\eps$ except at the boundary. Within the Ekman boundary layer, Figure \ref{fig:criticalprofiles} (second and fourth rows) shows that the numerically computed full problem (open circles), the outer solution (dashed-dotted line) and the composite problem from the superposition of inner and outer solutions (solid line). The composite solutions appear to  match quantitatively at leading order. 
Note that for no-slip case, $u^{(o)}$, $\zeta^{(o)}$, and $w^{(o)}$ are all non-zero on the boundary, but the composite solution correctly captures the decay to zero. The same is true for $w^{(o)}$ in the stress-free case.

\begin{figure}
    \centering
    \includegraphics[width = \linewidth]{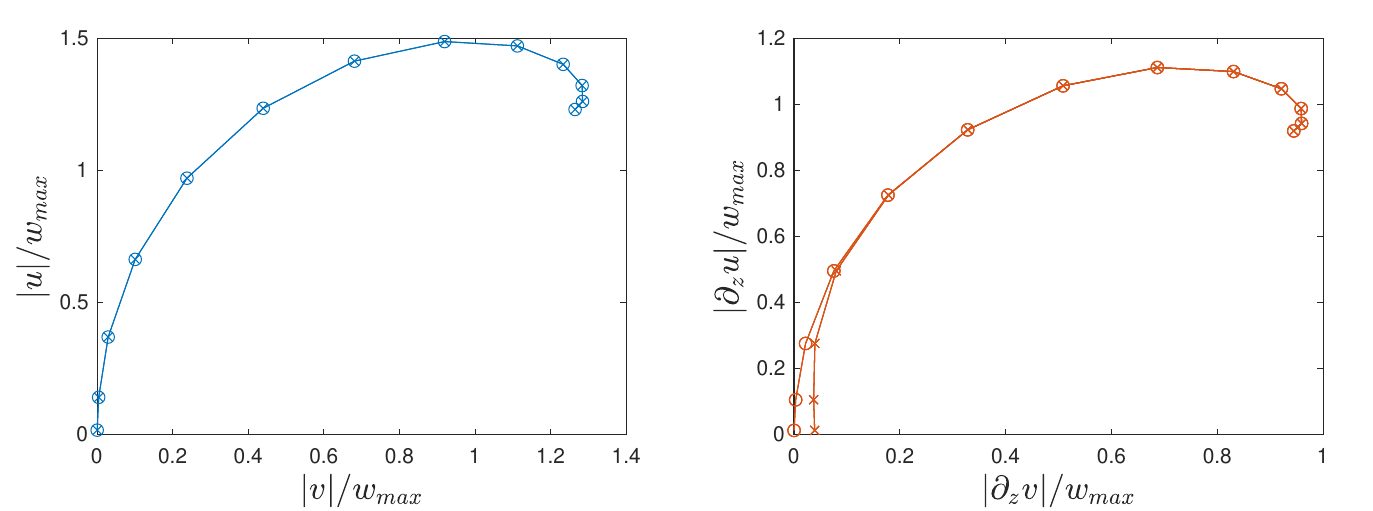}
    \caption{\small{Hodographs of the horizontal velocity in the Ekman layer for no-slip (left, blue) and stress-free (right, red) problems for $\eps = 10^{-4}$, $\vartheta_f = \pi/4$, $\chi = \pi/4$, and critical $\Ra$ and wavenumber. For the no-slip case, $u$ is plotted against $v$, but for stress-free we show $\pd{z}u$ versus $\pd{z}v$. The open circles denote the numerically computed full problem, and the $\times$'s are the composite solution (the numerically computed interior plus the analytic boundary layer). }}
    \label{fig:spirals}
\end{figure}

In the lower row of Figure \ref{fig:criticalprofiles}, we plot the profiles for $\haz\cdot\nabla (u,\zeta)$  in the stress-free case instead of just ($u$, $\zeta$), since this is the quantity used to set the pumping condition. In the immediate vicinity of the boundary, this shows  slight differences between the solution with pumping and the exact stress-free boundary conditions that were not apparent if the derivative is not plotted.
Figure \ref{fig:spirals} also illustrates this results as a hodograph of $\pd{z} u$ versus $\pd{z} v$ (the no-slip result $u$ versus $v$ is also included in the left plot). For the stress-free case, there is an observable $\mathcal{O}(\eps^{1/2})$ error in $\pd{z}(u,v)$ between the full and composite solutions due to the fact that the boundary condition (\ref{eq:noslipandstressfreeBCs}b) is only satisfied to leading order in the composite solution. This may be understood as follows. Recall, a stress-free boundary requires 
\begin{equation}
    \begin{split}
         \haz \cdot \nabla \boldsymbol{v}\equiv \lb -\gamma \pd{y} +\eps\pd{\Omega} \rb \boldsymbol{v}^{(o)} +
          \lb -\gamma \pd{y} + \delta \eps^{-1/2} \pd{\mu}  \rb \boldsymbol{v}^{(i)} =0.
    \end{split}
\end{equation}
However, pumping boundary conditions are deduced from the leading order expression
\begin{equation}
    \begin{split}
        \lb -\gamma \pd{y} +\eps\pd{\Omega} \rb \boldsymbol{v}^{(o)} +
           \delta \eps^{-1/2} \pd{\mu}   \boldsymbol{v}^{(i)} =0
    \end{split}
\end{equation}
the difference being $\mathcal{O}(\gamma \pd{y}\boldsymbol{v}^{(i)} ) = \mathcal{O}(\eps^{1/2}) $ given that 
$\boldsymbol{v}^{(i)}=\mathcal{O}(\eps^{1/2})$. 

\section{Strongly Nonlinear Solutions}
\label{sec:singlemode}
\subsection{The QG-RBC model}
The QG-RBC equations (\ref{eq:QGRBC}) admit exact steady nonlinear single-mode solutions of the form
\begin{subequations}
\label{eqn:smb}
\beginar
& \lb w_0, \theta_1\rb =(\widehat w,\widehat \theta)(\Omega) h(x,y) + c.c.,\quad 
\psi_0 =  \displaystyle{\lb \widehat \psi(\Omega) h(x,y) + \frac{\gamma}{k_\perp^2} \widehat w(\Omega) h_x(x,y) \rb + c.c.} \hspace{3em}
\endar
with 
\beginar
\widehat \theta(\Omega) =\displaystyle{ -\frac{ \sigma}{k_\nabla^2}   \lb \pd{\Omega}\overline{\Theta}_0 -1 \rb \widehat w(\Omega)}.
\endar
\end{subequations}
Here $h(x,y)$ satisfies the planform equation $\nabla^2_\perp h = - {k}_{\nabla}^2 h $. Single-mode solutions require the Jacobian advection terms in the QG-RBC \eqref{eq:QGRBC} be identically zero under this ansatz. We note, the only such solutions for $\gamma\ne0$ are roll solutions $h(x,y)=\exp\left( ik_x x+ik_y y\right)$. Single-mode solutions are known to be unstable to fully 3D multimodal perturbations \citep{mS06}, however, they provide a skeletal framework for dynamical trajectories within phase-space and thus highly influence the evolution of realized solutions. Here again, the dependence of $\psi$ within the expression for vertical motions $\widehat w(\Omega)$ is a reflection of the  non-axial buoyant driving of axial vorticity. The amplitudes $\widehat w(\Omega), \widehat \psi(\Omega)$  and mean temperature gradient  $ \pd{\Omega}  \overline{\Theta}-1$ satisfy the coupled ODE system
\begin{subequations}
\label{eqn:smodeb}
\beginar
& \pd{\Omega} \widehat w + k_\perp^2  k_\nabla^2 \widehat \psi =0 , \qquad 
\pd{\Omega} \widehat \psi -  \lb \displaystyle{ \frac{\Ra Nu }{ k_\nabla^2 + 2\sigma^2
\vert\widehat w\vert^2} } -  \frac{ k_\nabla^4}{ k_\perp^2}
 \rb \widehat w =0 ,\\
& \pd{\Omega} \overline{\Theta}_0 -1 = \displaystyle{ - \frac{k_\nabla^2 Nu }{k_\nabla^2 + 2\sigma^2 
\vert\widehat w\vert^2} }
\endar
with Nusselt number  
\beginar
Nu = \lsq \int_0^1 \frac{k_\nabla^2}{k_\nabla^2 + 2\sigma^2
\vert\widehat w\vert^2} d \Omega \rsq^{-1} 
\endar
\end{subequations}
 measuring the nondimensional heat transport.  Without loss of generality, the dependency of $\sigma$ 
 can be absorbed by rescaling amplitudes according to $(w_0,\psi_0,\theta_1)\mapsto (w_0,\psi_0,\theta_1)/\sigma$.
 System \eqref{eqn:smodeb} is accompanied with impenetrable boundary conditions $\widehat w(0)=\widehat w(1) =0$. Equations (\ref{eqn:smodeb}a) may also be collapsed to
 \beginar
 \label{eqn:csme}
 \pd{\Omega\Omega} \hat w + k^2_\perp k^2_\nabla \lb \frac{\Ra Nu}{k^2_\nabla + 2 \sigma^2 \vert \hat w\vert^2} - \frac{k_\nabla^4}{k_\perp^2}\rb \hat w = 0, \qquad w(0)=\widehat w(1) =0. 
 \endar
The single-mode analysis of \citet{iG15} for the upright QG-RBC may be extended to the tilted $f$-plane for both the QG-RBC with impenetrable boundaries and CQG-RBC with stress-free pumping conditions. Ensuring that all terms in \eqref{eqn:csme} are dominant at the midplane $\Omega=0.5$ requires
\beginar
\vert \hat w(0.5) \vert^2 \sim \frac{k^2_\perp}{k^4_\nabla} \Ra Nu, \quad\implies\quad 
\pd{\Omega}\overline{\Theta}(0.5) - 1\sim \frac{k^6_\nabla}{k^2_\perp}\frac{1}{\Ra}, \qquad \epsilon > 0.
\endar
The latter result follows from (\ref{eqn:smodeb}b). Along loci $k_\perp \propto \Ra^\alpha$ 
 \citet[][equations (33) \& (41)]{iG15} has established the sharp bounds
 \beginar
 \Ra^{1+2\alpha} \le &Nu& \le  \Ra^{1+2\alpha}\ln \lb  \Ra^{1-4\alpha} Nu\rb \sim  \Ra^{1+2\alpha + \epsilon}\endar
 for $-1/2\le\alpha\le 1/4$ in QG-RBC, which imply
 \beginar
 \pd{\Omega}\overline{\Theta}(0.5) - 1&\sim& \Ra^{4\alpha -1}.
 \endar     
 We note that the analysis for the CQG-RBC with no-slip pumping boundary conditions remains an open problem. Thus, as points of reference for comparison the  QG-RBC (or CQG-RBC with stress-free boundary conditions) give 
 \beginar
 \label{eqn:scalep}
\begin{array}{lcl}
  k_\perp = \mathrm{fixed},\ \ \alpha=0, & \implies &\pd{\Omega}\overline{\Theta}\sim \Ra^{-1},
 \\ \\
  k_\perp = \Ra^{1/4}, & \implies & \pd{\Omega}\overline{\Theta}\sim \mathrm{const.}
\end{array}
\endar
The scaling exponent 
$\alpha=1/4$ corresponds to the maximal heat transport for the single-mode solutions in QG-RBC and provides the upper bound $Nu\sim\Ra^{3/2}$ compared to the fixed wavenumber case where $Nu\sim\Ra$.

\subsection{The CQG-RBC model}
Single-mode solutions to the CQG-RBC model of the form \eqref{eqn:smb} may be pursued upon neglecting nonlinear vertical advection of temperature fluctuation in (\ref{eq:CQGRBC}c) that are significant in the thermal wind layer. This results in 
complex-valued system
\begin{subequations}
\label{eqn:smodec}
\beginar
 \pd{\Omega} \widehat w + k_\perp^4 \lb 1+ \gamma^2 \sin^2 \chi \rb \widehat \psi &=& 0 , \\ 
 \pd{\Omega} \widehat \psi +  k_\perp^2 \lb 1+ \gamma^2 \sin^2 \chi \rb^2 \widehat w &=& \displaystyle{\frac{\Ra}{\sigma}} \widehat\theta,\\
\sigma  \lb \pd{\Omega} \overline{\Theta}_0 - 1 \rb \widehat w =  \lsq  \eps^2 \pd{\Omega\Omega} - k_\perp^2 \lb 1+ \gamma^2 \sin^2 \chi \rb  \right.  &-& \left. 2 i  \eps \gamma k_\perp \sin\chi \pd{\Omega} \rsq \widehat \theta,  \\
\sigma \pd{\Omega} \lb \widehat w \widehat \theta^* + c.c. \rb  &=& \pd{\Omega\Omega} \overline{\Theta}_0.
 \endar
\end{subequations}
 System \eqref{eqn:smodec} is accompanied with fixed-temperature conditions  $\widehat \theta(0)=\widehat \theta(1)=0$,
 and no-slip pumping boundary conditions given in \eqref{eq:pumpingconditionsw}. Recall, for stress-free  pumping boundary conditions the CQG-RBC is equivalent to the QG-RBC.

\begin{figure}
    \centering
    \includegraphics[width = \linewidth]{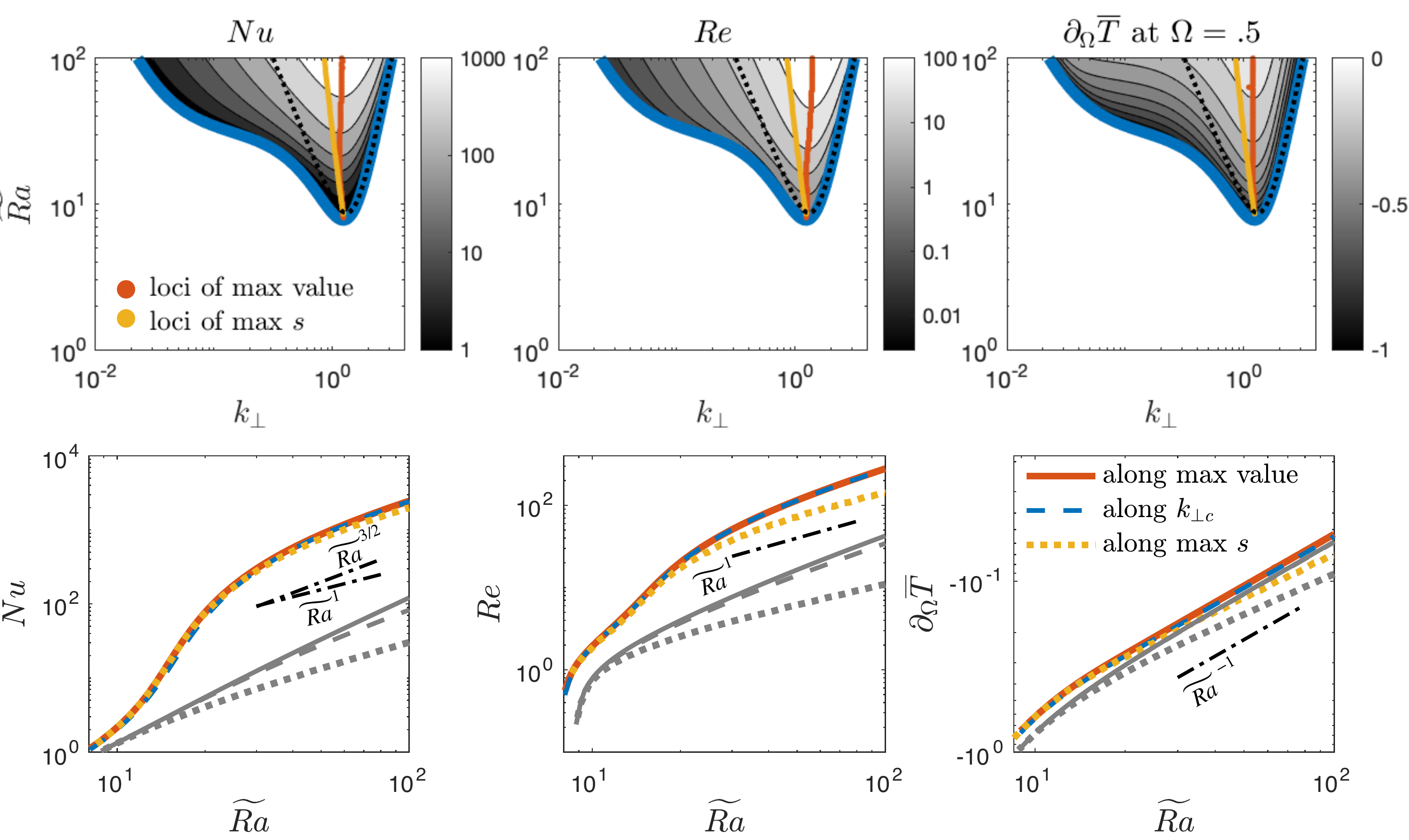}
    \caption{\small{The CQG-RBC model with no-slip pumping boundary conditions evaluated on 
    tilted $f$-plane for north-south rolls at any arbitrary $\vartheta_f< 90^\circ$ with $\eps = 10^{-2}$. Top Row: Contours of $Nu$, $Re$, and $\partial_\Omega \overline{T}$ at the midplane in the $k_\perp$ vs $\Ra$ plane. The solid blue denotes the marginal stability curve. For comparison, the dotted black line is the analytic QG marginal stability curve where pumping is omitted. Bottom Row: Plots of $Nu$, $Re$, and $\partial_\Omega \overline{T}$ as a function of $\Ra$ along loci for maximal values at each $\Ra$ (red solid line), maximal linear growth rate $s$ (yellow curve), and fixed critical wavenumber $k_{\perp c}$ (blue dotted line).  For comparison, results for the QG model along identical loci are illustrated (grey lines). }}
    \label{fig:NuReDT_contours_upright}
\end{figure}

\begin{figure}
    \centering
\includegraphics[width = \linewidth]{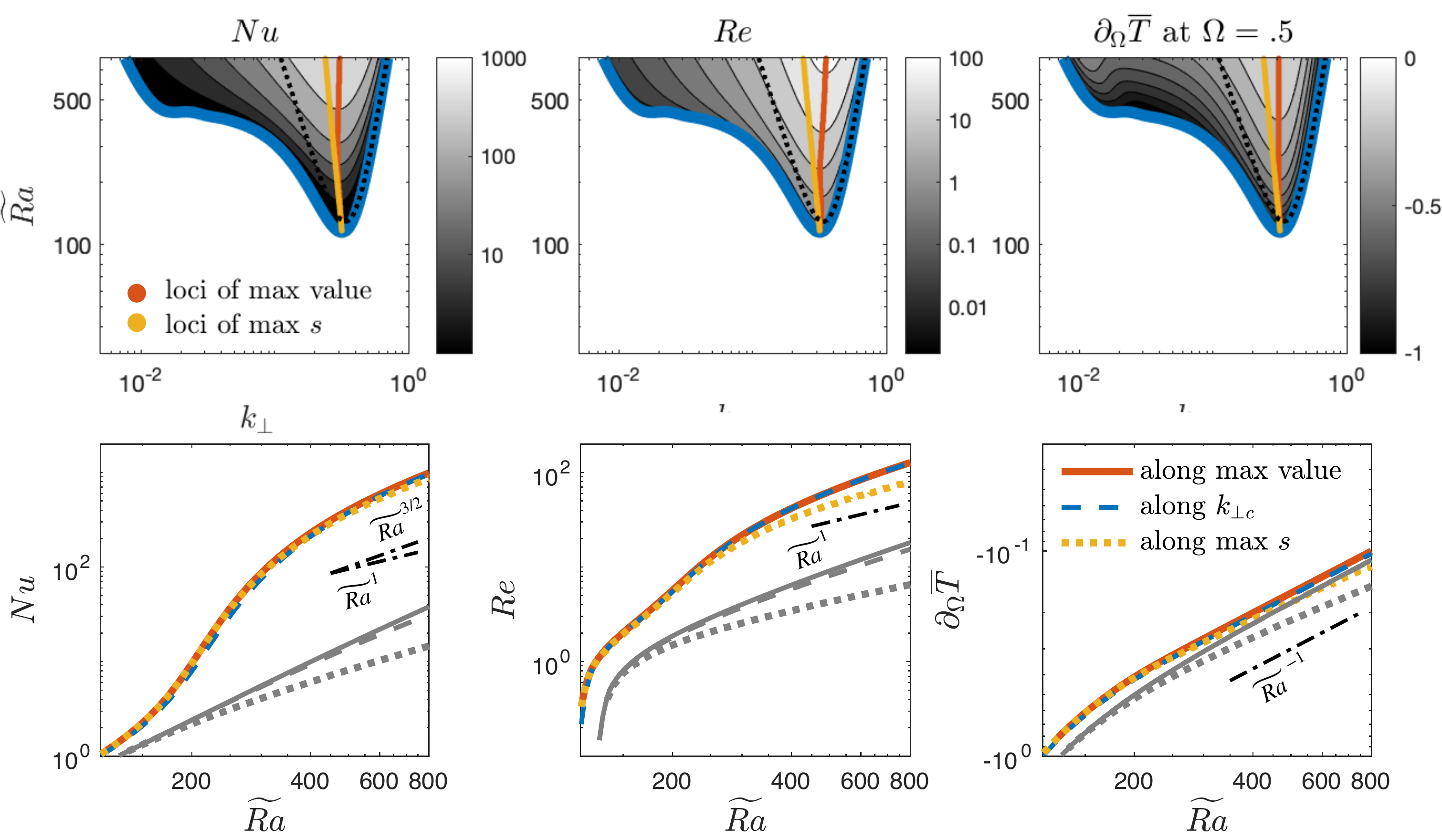}
    \caption{\small{The CQG-RBC model with no-slip pumping boundary conditions evaluated on the tilted $f$-plane at $\vartheta_f=75^\circ$ for east-west rolls ($\chi = 90^\circ$) with $\eps = 10^{-2}$. Labeling as in Figure~\ref{fig:NuReDT_contours_upright}.  }}
    \label{fig:NuReDT_contours_tilted}
\end{figure}

\subsection{Results: Fully Nonlinear Single-mode Solutions} 
Investigations of single-mode solutions to the QG-RBC model (equation \eqref{eqn:smodeb}) have been performed by \citet{iG15} and  \citet{kJ98} in the absence of Ekman pumping for the upright and tilted $f$-plane cases respectively. \citet{jCKMSV16} explored the impact of pumping boundary conditions for upright RRBC.
Such solutions are asymptotically accurate but unstable solutions to the rapidly rotating RBC problem in the limit $\varepsilon\rightarrow 0$. In totality, at a fixed $\Ra$ these solutions may be interpreted as the skeletal structure of the high-dimensional phase space that all realized solution trajectories must navigate. Hence, under the assumption that they possess a close  proximity to the realized fluid state, the global properties of single-mode solutions are informative. Here, fully nonlinear single-mode roll solutions are investigated in the $\Ra$ vs $k_\perp$ plane via a simulation suite of the CQG-RBC model (equation \eqref{eqn:smodec}) with no-slip pumping boundary conditions at arbitrary $\vartheta_f$. This is compared with an identical simulation suite for the QG-RBC model that has been established as equivalent to the CQG-RBC model with stress-free pumping conditions. 

In figure \ref{fig:NuReDT_contours_upright}, contour plots are illustrated at $\varepsilon=10^{-2}$ for $Nu$, $Re = \max(\vert \widehat w\vert)$, and midplane mean temperature gradient $\pd{\Omega}\overline{T} = \pd{\Omega}\overline{\Theta}_0-1$ obtained from the CQG-RBC model on the tilted $f$-plane for north-south rolls at any arbitrary $\vartheta_f< 90^\circ$, plots (a)-(c). We recall, based on the co-latitudinal Rayleigh number $\Ra$, north-south rolls have the same linear and nonlinear stability properties at any given $\vartheta_f$. 
As a function of $\Ra$ at fixed $k_\perp$ it can be observed from the contours that $Nu$ and $Re$ are monotonically increasing functions of $\Ra$ while $\pd{\Omega}\overline{T}$ is a monotonically decreasing function of $\Ra$ indicating approach to an isothermal interior. In figure \ref{fig:NuReDT_contours_tilted}, the opposite bookend case for east-west rolls is illustrated at co-latitudinal location $\vartheta_f= 75^\circ$ and display identical features to the north-south case in figure \ref{fig:NuReDT_contours_upright} albeit for delayed $\Ra$ due to the increased stability of this roll orientation (see plots (a)-(c)).

Lineplots as a function of $\Ra$ along loci highlighted in (a)-(c) for fixed critical $k_{c\perp}$, maximal linear growth rate $\propto k^{-8}_\perp$, and extremal values $k_{max}$ achieved at a fixed $\Ra$ are given in plots (d)-(f) for $Nu$, $Re$, and $\pd{\Omega}\overline{T}$ respectively (colored lines). These loci are motivated by linear marginal onset, recent simulations of the QG-RBC that find that integral length scale of convection follows the maximal growth rate \citep{tO23}, and exploration of optimal global transport of heat and momentum. For comparison, results along the equivalent loci are given for the QG-RBC model that omits pumping (grey lines). As established, this is equivalent to the CQG-RBC with stress-free pumping conditions. It is evident that Ekman pumping in presence of no-slip boundaries significantly enhances the global heat and momentum transport as measured by Nusselt number  $Nu$ and Reynolds number $Re$. It is observed that the line-plots for $Nu$ in the CQG-RBC model are insensitive to the particular choice of locus. Thus, we consider fixed $k_\perp$ and $k_{max}$ as the representative markers,
and plot for guidance scaling lines $Nu\sim\Ra^1$ and $Nu\sim \Ra^{3/2}$, $Re = \vert \hat w(0.5)\vert \sim \Ra^1$ and $\pd{\Omega}\overline{\Theta}\sim Ra^{-1}$ established for the QG-RBC model (see equation \ref{eqn:scalep}). It can be seen that curves associated with loci of $k_\perp$ fixed and extrema for $Nu$, $Re$, and $\pd{\Omega}\overline{T}$  are in compliance with these estimated.  Thus the impact of Ekman pumping appears to reside in the prefactor, this is consistent with the findings of \citet{mP17}.  However, instantaneous scaling exponents along curves of maximal growth rate $k_\perp\propto \Ra^{-1/8}$ appear consistently smaller for $Re$ and $\pd{\Omega}\overline{T}$. This is an expected result given that the single-mode theory is one that  captures nonlinear stationary states and thus excludes consideration of maximal growth solutions. Consequently, loci tracking maximal values of contours at fixed $\Ra$ constitute upper bounds.

These qualitative features illustrated in figures~\ref{fig:NuReDT_contours_upright} and \ref{fig:NuReDT_contours_tilted} extend to cases with decreasing $\varepsilon$ across all co-latitudes $\vartheta_f < 90^\circ$ (see also \citet{jCKMSV16}). As with the linear results, $Nu$ and $Re$ vs $\Ra$ for the CQG-RBC model experiences a delayed departure from that observed in the QG-RBC model as $\varepsilon$ decreases. However, the transition to a power law scaling is increasingly abrupt as $\varepsilon$ decreases and the larger values of $Nu$ and $Re$ are observed as in \citet{jCKMSV16}.

\section{Discussion and Conclusion} 
\label{sec:discussion}
The boundary layer reduction of rapidly rotating Rayleigh-B\'enard convection is considered on the tilted $f$-plane located at arbitrary co-latitude $\vartheta_f< 90^\circ$. As a consequence of gyroscropic alignment occurring through the Taylor-Proudman constraint, spatial variations of fluid structures along the axis of rotation are observed to be $\mathcal{O}(H)$ as compared to $\mathcal{O}(\ek^{1/3}H)$ along $\haz$ (i.e. radial) direction. This motivates the use of a non-orthogonal coordinate system representation where the upright coordinate aligns with the rotation axis as opposed to gravity. A matched asymptotic analysis is performed on the incompressible Navier-Stokes equations (iNSE) that governs the fluid dynamics. Three regions are identified and matched asymptotically: a geostrophic interior whose velocity and thermal fields are respectively rectified by an inner Ekman boundary layer of $\mathcal{O}(\ek^{1/2} H)$ and a middle thermal wind layer of $\mathcal{O}(\ek^{1/3} H)$. The analysis reveals that these boundary layers obey classical equation sets but evolve with the boundary layer coordinates that align with the rotation axis. Specifically, an analysis of the Ekman layer yields the fourth-order ODE system resulting from the Coriolis-viscous force balance \citep{hG69}.  Mass continuity then uncovers parameterized boundary conditions which serves as the kinematic condition that circumvents the numerical spatial resolution requirements of a viscous layer and captures the effects of Ekman pumping and suction. Closure of the iNSE system that utilizes this kinematic condition requires it be supplemented with geostrophic boundary conditions serving as the mechanical boundary conditions for the interior dynamics. By contrast, for the non-hydrostatic quasi-geostrophic equations (i.e., QG-RBC and CQG-RBC) constituting the asymptotic reductions of the iNSE in the limit of rapid rotation, no mechanical boundary conditions are required. The thermal wind layers are in geostrophic and axial hydrostatic balance, the latter balance ensuring thermal fluctuations that maintain fixed temperature boundary conditions.

In the presence of no-slip boundaries, the parameterized boundary condition is the vertical velocity/vertical vorticity pumping relationship, $w\propto\varepsilon^{1/2} \zeta$, or in dimensional terms, $w^*\propto (\nu/2\textOmega\cos\vartheta_f)^{1/2} \zeta^*$. This is known in literature through its application to large-scale atmospheric and oceanic flows \citep{gV06}
but less familiar to convectively-driven flows. Linear stability investigations of the iNSE and the CQG-RBC with parameterized pumping boundary conditions reveal that they are a quantitatively accurate alternate to the unapproximated problem where Ekman boundary layers are unfiltered. Importantly, to our knowledge, it is demonstrated for the first time that Ekman pumping strongly destabilizes large-scale (low-wavenumber) convective modes  and  thus significantly extended the spatial range of convectively unstable modes at fixed $\Ra$. 
It is established that this occurs when pumping velocity $w\sim\mathcal{O}(1)$ which is always achieved in the quasi-geostrophic regime established to have the upper bound $\zeta=o(\varepsilon^{-1})$. This implies some caution should be taken not to truncate the dynamical regime in selecting the aspect ratio of computational domains in plane-layer investigations of  RRBC. 

For stress free boundary conditions, the asymptotic analysis uncovered the vertical velocity/vertical gradient of vertical vorticity pumping relationship, $w\propto\varepsilon {\hat \haz} \cdot \nabla  \zeta$, or in non-dimensional terms, $w^*\propto (\nu/2\textOmega\cos\vartheta_f) {\hat \haz} \cdot \nabla^* \zeta^*$. To our knowledge this result is not known in the literature. Linear stability theory of the iNSE and CQG-RBC with pumping boundary conditions again reveal excellent quantitative agreement with the iNSE without approximation. In fact, it is found that all three of these models are accurately captured by the QG-RBC  constrained only by the requirement of impenetrable boundary conditions. This is supported by the observation that the pumping velocity always remains subdominant in the quasi-geostrophic regime where  $\zeta=o(\varepsilon^{-1})$. Thus Ekman boundary layers while present remain passive. The pumping boundary conditions for this case thus serve solely as a means of filtering these layers thus providing relief on the numerical spatial resolution requirements. 

Results from DNS with imposed pumping conditions will be pursued in the future. As an intermediate step, results for single-mode solutions to the  CQG-RBC model were presented for both no-slip and stress-free boundary conditions. It is demonstrated that pumping in the presence of no-slip boundaries greatly enhance the global heat and momentum transport properties of the fluid layer to the remarkable extent that an $\mathcal{O}(\ek^{1/2} H)$ layer generates $\Delta Nu, \Delta Re = \mathcal{O}(1)$. We note that single-mode solution while instructive omit an important phenomena, i.e., the lateral stirring and mixing of thermal field. As such the mean temperature field does not saturate to an unstable profile as $\Ra\rightarrow\infty$ as observed in fully nonlinear simulations \citep{kJ96,mS06,kJ12}. Instead, it continues to an isothermal interior; $\pd{\Omega} \overline{T}\sim\Ra^{-1}$ for stress-free boundaries.
This feature is inherent to the single-mode approximation including recent works of \citet{aB14,barker2020}  based on the original work of \citet{dS79} that report mean temperature gradient power laws that evolve to isothermality.

\section*{Acknowledgements}
This work was supported by the National Science Foundation (Grant DMS-2308337). K.J. thanks Dr. Geoff Vasil for fruitful interactions and discussions and Dr. Jonathan Aurnou for useful remarks on the manuscript.

\appendix
\section{Mixed vorticity-velocity formulation}
\label{sec:mvort}
The primitive variable formulation of the linearized iNSE (\ref{eq:covariantbasisequationsrescaledfull}) in the main text is of $9^{th}$ order in $\Omega$. Specifically, the continuity equation requires the imposition of an $9^{th}$ auxiliary boundary condition applied to the pressure function $p$. Instead of pursuing this option, we numerically solve the following modified set of equations for the variables $\ub= u \hx + (v-\gamma w) \hy + w/\eta_3 \hz $, $\boldsymbol{U}_\perp=U^g\hx + V^g\hy$, $\boldsymbol{\omega}=\omega^1 \hx + \omega^2 \hy + \zeta/\eta_3 \hz$, $p$ and $\theta'$:
\begin{subequations}
\beginar
    \omega^1 &=&  \pd{y}\lb w+\gamma v\rb -\eps \pd{\Omega} v,\\ 
    \omega^2 &=& -\pd{x}\lb w+\gamma v\rb +\eps \pd{\Omega}u,\\
    \zeta &=&   \pd{x}v-\pd{y}u,
    \endar
\beginar
      U^g &=& \eps^{-1}\lb u+\pd{y}p\rb,\\
      V^g &=& \eps^{-1} \lb v-\gamma w \rb - \pd{x}p,\\
  \pd{t} u - V^g &=& \lb\eps \pd{\Omega}- \gamma\pd{y}\rb\omega^2+\lb\eps \gamma\pd{\Omega}-\frac{1}{\eta_3^2}\pd{y} \rb \zeta,\\
\pd{t} \lb v -\gamma w\rb + \frac{1}{\eta_3^2}U^g &=&  -\eps\pd{\Omega}\omega^1 + \gamma \pd{x} \omega^2+\frac{1}{\eta_3^2}\pd{x}\zeta +\gamma \pd{\Omega}p - \frac{\gamma \Ra}{ \sigma} \theta\ \ \ \\
   \pd{t}  w +\gamma U^g  & = & \frac{1}{\eta_3} \lb {\gamma}\pd{x}\zeta -\pd{y}\omega^1 + \pd{x}\omega^2\rb +\pd{\Omega}p +  \frac{\Ra }{ \sigma} \theta \\
   \pd{t} \theta  - w &=& \frac{1}{\sigma}\lb \pd{x}^2\theta+\frac{1}{\eta_3^2}\pd{y}^2\theta - 2\eps\gamma \pd{y}\pd{\Omega} \theta +\eps^2 \pd{\Omega}^2 \theta\rb,\\
 \pd{x} U^g +\pd{y} V^g +\pd{\Omega}w  &=& 0.
\endar
\label{eq:linearcovariantfullderivsteady}
\end{subequations}
This is a closed formulation that remains $8^{th}$ order, i.e.,  compatible with either the number of physical or pumping boundary conditions presented in Table~\ref{Table:EquationSets}.

\subsection{Boundary conditions} In the mixed vorticity-velocity formulation, we may avoid setting boundary conditions on the mixed derivative $\haz \cdot \nabla$ by using the following identities. For the unapproximated stress-free iNSE problem,
\begin{subequations}
\beginar
    \haz \cdot \nabla u = \omega^2 +\gamma \zeta \quad \mbox{on}\ \Omega &=& 0,1,\\
     \haz \cdot \nabla v = \omega^1\quad  \mbox{on}\ \Omega &=& 0,1,
    \endar
\end{subequations}
so the mixed boundary conditions on $u$ and $v$ become Dirichlet conditions on $\omega^1$ and $\omega^2+\gamma \zeta$.
For the parameterized iNSE stress-free problem, 
\begin{equation}
     \haz \cdot \nabla \zeta = -\lb \pd{x} \omega^1 + \pd{y}\lb \omega^2 +\gamma \zeta \rb\rb,
\end{equation}
so the pumping boundary condition (\ref{eq:pumpingconditionsw} b) can be formulated as a Dirichlet condition on $\omega^1$, $\omega^2$, and $\zeta$.

\section{Numerics}
\label{sec:NumA}
\subsection{Linear problems}
For the linear stability problem, we assume solutions of the form
\begin{equation}
    \begin{pmatrix} 
\omega^1 \\ \omega^2 \\ \zeta \\ 
u \\ v\\ w \\ p\\ \theta
    \end{pmatrix} = \begin{pmatrix} 
\hat{\omega}^1 \\ \hat{\omega}^2 \\ \hat{\zeta} \\ 
\hat{u} \\ \hat{v}\\ \hat{w}\\ \hat{p}\\ \hat{\theta}
    \end{pmatrix}\exp\lb st + ik_x x +ik_y y\rb.
\end{equation}
We expand the fluid variables in a recombined Chebyshev basis which makes applying boundary conditions sparse. For almost all variables, we use the expansion
\begin{equation}
    \sum_{n=0}^{N-1}c_{(n)}\varphi_n(r),
    \label{eq:generalgalerkinexpansion}
\end{equation}
where $r = 2\Omega -1 \in [-1,1]$, and $\varphi_n$ is a Dirichlet bases \citep{Ded2020}, i.e., a  Chebyshev Galerkin polynomials given by
\begin{equation}
\begin{split}
    \varphi_0(r) = T_0(r) = 1, \quad \varphi_1(r) = T_1(r) = r,\quad  \varphi_n(r) = T_n(r)-T_{n-2}(r) \quad \mbox{for}\ n\geq 2,
\end{split}
\end{equation}
where $T_n(r) = \cos(n\arccos(r))$ are the standard Chebyshev polynomials. Then $\varphi_n(\pm 1) = 0$ for $ n\geq 2$, so all that is required to enforce a Dirichlet boundary condition at the top and bottom are the equations 
\begin{equation}
    c_{(0)} \pm c_{(1)} = 0,
\end{equation}
independent of $N$. For the stress-free pumping boundary conditions we require a mixed derivative $\haz \cdot \nabla$, so we use the basis 
\begin{subequations}
    \beginar
        \psi_m &=& T_m,  \quad m = 0,1,2,3\\
        \psi_m &=& T_{m-4} - \frac{2(m-2)}{m-1}\ T_{m-2}+\frac{m-3}{m-1}\ T_m, \quad m = 4,5,...,N-1
    \endar
\end{subequations}
\citep{kJ09}. In this basis, $\psi_m(\pm 1) = \psi_m'(\pm 1) = 0$ for $m\geq 4$, so 
\begin{equation}
\begin{split}
   \lb \haz \cdot \nabla\rb &\sum_{m=0}^{N-1} c_{(m)}\psi_m\biggr\rvert_{r\ =\  \pm 1} = \sum_{m=0}^{N-1} c_{(m)}\lb 2\eps \pd{r}-ik_y \gamma \rb \psi_m\biggr\rvert_{r\ =\  \pm 1}\\
   &=2 \eps \lb c_{(0)}\pm  c_{(1)}+ c_{(2)}\pm c_{(3)} \rb -ik_y \gamma \lb  c_{(1)}\pm 4  c_{(2)} +9 c_{(3)} \rb,
\end{split}
\end{equation}
so enforcing this boundary condition is also independent of $N$. We also employ a quasi-inverse technique to treat the vertical derivatives.

\subsection{Nonlinear problems}
We solve the nonlinear singlemode problem using MATLAB's bvp5c. This requires a first-order formulation. In order to ensure real-valued variables, we write the roll ansatz as sines and cosines, 
\begin{subequations}
    \beginar
 w_0 &=&  w_c \cos(k_x x+k_y y) + w_s \sin(k_x x +k_y y)\\
 \theta_1 &=&  \theta_c \cos(k_x x+k_y y) + \theta_s \sin(k_x x +k_y y)\\
 \psi_0 &=& \psi_c \cos(k_x x+k_y y) + \psi_s \sin(k_x x +k_y y )+\frac{\gamma}{k_x^2+k_y^2} \pd{x} w_0
    \endar \label{eq:sinecosinesinglemodeansatz}
\end{subequations}
substituted into the CQG-RBC (\ref{eq:CQGRBC}) yields the real-valued system
\begin{subequations}
    \beginar
\pd{\Omega}w_s &=& -k_p^2(k_x^2+k_y^2)\psi_s\\
\pd{\Omega}w_c &=& -k_p^2(k_x^2+k_y^2)\psi_c\\
\pd{\Omega}\psi_s &=& \frac{\Ra}{\sigma} \theta_s - \frac{k_p^4}{k_x^2+k_y^2} w_s\\
\pd{\Omega}\psi_c &=& \frac{\Ra}{\sigma} \theta_c - \frac{k_p^4}{k_x^2+k_y^2} w_c,\\
  \frac{\eps^2}{\sigma} \pd{\Omega}^2 \theta_s &=& \sigma w_s \lb  w_c\theta_c + w_s \theta_s \rb - Nu\ w_s +\frac{k_p^2}{\sigma} \theta_s - \frac{2 \eps \gamma k_y}{\sigma} \pd{\Omega}\theta_c\\
    \frac{\eps^2}{\sigma} \pd{\Omega}^2 \theta_c &=& \sigma w_c \lb  w_c\theta_c + w_s \theta_s \rb - Nu\ w_c +\frac{k_p^2}{\sigma} \theta_c + \frac{2 \eps \gamma k_y}{\sigma} \pd{\Omega}\theta_s
    \endar
    \label{eq:sinecosineasnsatzfirstordersinglemode}
\end{subequations}

\bibliographystyle{jfm}
\bibliography{ref}

\end{document}